\documentclass[sigconf]{acmart}

\usepackage[english]{babel}
\renewcommand\footnotetextcopyrightpermission[1]{} 
\acmDOI{}
\acmISBN{}
\acmPrice{}
\usepackage[english]{babel}

\usepackage{epsfig}
\usepackage{balance}
\usepackage{pifont}

\usepackage{times}
\usepackage{graphicx} 
\usepackage{subcaption}
\usepackage{tabularx}
\usepackage{xspace}
\usepackage{listings}
\usepackage{verbatim}
\usepackage{booktabs}
\usepackage{colortbl}
\usepackage{nicefrac}
\usepackage{siunitx}
\usepackage{stackengine}
\usepackage[small,compact]{titlesec}
\titlespacing*{\section}{0pt}{4pt}{4pt}
\titlespacing*{\subsection}{0pt}{4pt}{4pt}
\titlespacing*{\subsubsection}{0pt}{0pt}{0pt}
\usepackage{amsmath}
\usepackage[subtle]{savetrees}

\usepackage{color}
\definecolor{darkred}{rgb}{0.7,0,0}
\definecolor{darkgreen}{rgb}{0,0.5,0}
\hypersetup{colorlinks=true,
        linkcolor=darkred,
        citecolor=darkgreen}

\newcommand{\eg}{{e.g.,} }

\newcommand{\ie}{{i.e.}, }

\newcommand{\Fig}[1]{Fig.~\ref{fig:#1}\xspace}

\newcommand{\Sec}[1]{$\S$\ref{s:#1}\xspace}
\newcommand{\App}[1]{Appendix~\ref{app:#1}\xspace}

\newcommand{\pg}[1] {{\textcolor{red}{PG: #1}}}
\newcommand{\an}[1] {{\textcolor{blue}{AN: {#1}}}}

\newcommand{\cut}[1]{}

\newcommand{\ct}{\small \tt}

\newcommand{\dcc}{delay-controlling\xspace}

\newcommand{\rtt}{round-trip time\xspace}

\newcommand{\RTT}{RTT\xspace}

\newcommand{\pulser}{pulser\xspace}

\newcommand{\watcher}{watcher\xspace}
\newcommand{\watchers}{watchers\xspace}

\newcommand{\mahimahi}{Mahimahi\xspace}
\newcommand{\name}{Nimbus\xspace}

\newcommand{\possname}{Nimbus's\xspace}

\def\compactify{\itemsep=0pt \topsep=0pt \partopsep=0pt \parsep=0pt}
\let\latexusecounter=\usecounter

\newenvironment{CompactEnumerate}
  {\def\usecounter{\compactify\leftmargin=17pt\latexusecounter}
   \begin{enumerate}}
  {\end{enumerate}\let\usecounter=\latexusecounter}

\titleformat{\section}
  {\normalfont\Large\bfseries}{\thesection}{1em}{\MakeUppercase} 

\begin{document}

\title{\LARGE Elasticity Detection: A Building Block for Internet Congestion Control\vspace{-2mm}}
\author{Prateesh Goyal$^{1}$, Akshay Narayan$^{1}$, Frank Cangialosi$^{1}$, Srinivas Narayana$^{2}$,\\Mohammad Alizadeh$^{1}$, Hari Balakrishnan$^{1}$\\\vspace{3mm}
{\large $^{1}$MIT CSAIL, $^{2}$Rutgers University}\vspace{2mm}}
\date{\vspace{5mm}}
%

\begin{abstract}
    This paper introduces Nimbus, a robust technique to detect whether the cross traffic competing with a flow is ``elastic'', and shows that this elasticity detector improves congestion control. If cross traffic is inelastic, then a sender can control queueing delays while achieving high throughput, but in the presence of elastic traffic, it may lose throughput if it attempts to control packet delay.  To estimate elasticity, Nimbus modulates the flow's sending rate with sinusoidal pulses that create small traffic fluctuations at the bottleneck link, and measures the frequency response of the rate of the cross traffic.  Our results on emulated and real-world paths show that congestion control using elasticity detection  achieves throughput comparable to Cubic, but with delays that are 50--70 ms lower when cross traffic is inelastic. \name detects the nature of the cross traffic more accurately than Copa, and is usable as a building block by other end-to-end algorithms.

\end{abstract}
\maketitle
\begin{sloppypar}

\section{Introduction}
\label{s:intro}


Achieving high throughput and low delay has been a key goal of congestion control research for decades. To achieve these goals, researchers have proposed many {\em delay-controlling} algorithms. These schemes (e.g., Vegas~\cite{vegas}, FAST~\cite{fasttcp}, LEDBAT~\cite{ledbat}, Sprout~\cite{sprout}, Copa~\cite{copa}) reduce their rates as delays increase to control packet delays and avoid ``bufferbloat''~\cite{bufferbloat}, unlike methods like Cubic~\cite{cubic}, NewReno~\cite{newreno}, and Compound~\cite{compound2} that must fill buffers to elicit congestion signals (packet losses or ECN).

There is, however, a major obstacle to deploying delay-controlling
algorithms on the Internet: their throughput is dismal when competing
against buffer-filling flows at a shared bottleneck. The reason
is that buffer-filling senders steadily increase their rates, causing queuing delays to rise; in response to increasing delays, a competing delay-controlling flow will reduce its rate. The buffer-filling flow then grabs this freed-up bandwidth. The throughput of the delay-controlling flow plummets, but delays don't reduce. Because most traffic on the Internet today uses buffer-filling algorithms, it is hard to justify deploying a delay-controlling scheme.

Is it possible to achieve the benefits of \dcc algorithms while ensuring that throughput does not degrade in the presence of buffer-filling schemes? We believe that a rigorous answer to this question requires the sender to understand the {\em nature} of the cross traffic. The salient aspect of this nature is whether the cross traffic is buffer-filling or not.  That, however, is beyond our current abilities, but  we contribute in this paper a new, rigorous algorithm to characterize whether cross traffic is {\em elastic} or not. We also show that elasticity is a good metric for determining whether the sender should attempt to control delays.

We define a flow to be elastic at a given bottleneck if it {\em increases its rate when it senses that more bandwidth is available there, and decreases it otherwise}. All other flows are {\em inelastic}. Examples of elastic flows include backlogged flows using either buffer-filling schemes like Cubic and NewReno, or delay-based schemes like Vegas, Copa, and BBR~\cite{bbr}. By contrast, constant bit-rate (CBR) flows, short TCP connections, application-limited flows, and flows bottlenecked at a different link are all inelastic. In general, traffic could have both elastic and inelastic flows; we define traffic to be elastic if it contains any elastic flows, and inelastic otherwise.



We have developed an {\em elasticity detector} called {\em Nimbus}, which any sender can use to make its congestion-control decisions. When Nimbus deems cross traffic to be inelastic, the sender can use a \dcc algorithm to achieve low delays, but when cross traffic is elastic, an algorithm that competes fairly with the cross traffic without necessarily attempting to control delay is required. 

Because all buffer-filling flows are elastic, this approach guarantees that a sender using Nimbus will not lose throughput by using a \dcc method when competing with such flows. It may, however, miss out on opportunities to control delays when cross traffic is both elastic and itself \dcc{} (note that this is really no different from the status quo, where a Cubic, BBR, or even Copa flow will crush Vegas, for example). Competing fairly with \dcc cross traffic while achieving low delay requires a method to determine not only the nature, but also the type of congestion control algorithm used by the cross traffic. 


\smallskip
\noindent
{\bf Elasticity detection.} Nimbus uses only end-to-end RTT measurements to monitor the cross traffic to determine if the cross traffic is elastic. The sender continuously modulates its rate with sinusoidal pulses to create small traffic fluctuations at the bottleneck at a specific frequency (e.g., 5 Hz). It concurrently estimates the rate of the cross traffic based on the its own send and receive rates, and monitors its frequency response (FFT) to determine if the cross traffic's rate oscillates at the same frequency. If it does, then the sender concludes that the cross traffic contains elastic flows; otherwise, it is inelastic. 

This technique relies on two assumptions. First, the sender must be able to create sufficient pulses
and observe the impact on cross traffic over a period of time. Thus it is best suited for large data transfers.
Fortunately, it is for such transfers that delay-controlling schemes are useful, because short flows are unlikely to cause significant queueing delay~\cite{bufferbloat}. 

Second, pulsing is most effective when the elastic flows react on a timescale of a few RTTs. 
If an elastic flow is slower to react, it can go undetected with short pulses. On the other hand, using longer pulses to detect such ``sluggish'' elastic flows could cause congestion. The majority of traffic on the Internet reacts on RTT timescales (e.g., ACK-clocked TCP flows). 
Nimbus is targeted at detecting ACK-clocked flows, but we have found that it also correctly classifies fast-reacting rate-based flows as elastic.

In our experiments, we find that \name is robust to a variety of cross traffic conditions, achieving at least 85\% detection accuracy even when  cross traffic is a combination of varying number of elastic flows and highly-varying inelastic short flows, or when cross traffic is composed of multiple elastic flows with different RTTs. These results hold across a wide range of network characteristics:  buffer sizes, RTTs, bottleneck link rates, active queue management schemes, and fraction of traffic controlled by Nimbus.

\smallskip
\noindent {\bf NimbusCC} is a congestion control system that uses elasticity detection to switch between TCP-competitive and delay-controlling modes. NimbusCC can support various algorithms in each mode. We report results with Vegas, Copa's default mode, and a simple new method that uses our cross-traffic rate estimator, as examples of \dcc algorithms, and Cubic and Reno as examples of TCP-competitive algorithms. 

\smallskip
\noindent {\bf Key results:} We have implemented \name{}CC in Linux using CCP~\cite{ccp-sigcomm18}. Our experimental results show that:
\begin{CompactEnumerate}
\item \name{}CC achieves throughput within 10\% of the fair share against elastic traffic made up of a variable number of TCP flows, whereas Copa is 54\% lower. NimbusCC also achieves 60~ms lower mean delay than Cubic against Poisson-distributed inelastic cross traffic. 

\item When cross traffic is modeled from a flow-size distribution measured at a WAN link~\cite{caida-dataset}, \name{}CC achieves throughput comparable to Cubic and BBR, but with 50~ms lower median delay. Copa has similar median throughput, but its 10th percentile of throughput is 60\% lower than fair share because it performs a lot worse than NimbusCC against elastic cross traffic. By contrast, both NimbusCC and Cubic (which NimbusCC emulates) have a 10th percentile only 30\% lower than fair share.

\item On 25 different Internet paths, \name{}CC achieved a throughput at least as high as Cubic, exceeding Cubic on paths with policers, with lower delays on 60\% of the paths and similar delays on the other 40\%. Compared to BBR, \name{}CC's throughput was 10\% lower, but the mean packet delay was 40--50 ms lower.
\end{CompactEnumerate}

\smallskip
\noindent
Our principal contribution is the idea that the nature of cross traffic, quantified as elasticity, is a useful building block for congestion control. We envision it being used to solve other problems in the future. For example, in tools like speedtest or iperf to inform users not only of the rate, but also whether the rate is lower than expected due to elastic cross traffic. Such a tool can shed light on traffic behavior and may also help guide the deployment of active queue management (AQM) schemes.  


\newcommand{\relworksec}{Related Work}
\section{\relworksec}
\label{s:related}


Copa~\cite{copa} aims to maintain a bounded number of packets in the bottleneck queue. Copa induces a periodic pattern of sending rate that nearly empties the queue once every 5 RTTs. This helps Copa flows obtain an accurate estimate of the minimum RTT and the queuing delay. In addition, Copa uses this pattern to detect the presence of  non-Copa flows: Copa expects the queue to be nearly empty at least once every 5 RTTs, provided only Copa flows with similar RTTs share the bottleneck link.
If the estimated queuing delay does not drop below a threshold in 5 RTTs, Copa switches to a TCP-competitive mode.

Unlike Copa, \name{} does not look for a pattern in the RTTs caused by its transmission pattern. Instead, it estimates the rate of the cross traffic and observes how the cross traffic reacts to the rate fluctuations it induces over a period of time. Thus, \name{} directly estimates the elasticity of cross traffic. 
Although elasticity detection takes a few seconds, our experiments show that it is more robust than Copa's method. We show that:
\begin{CompactEnumerate}
\item Copa incorrectly and frequently switches modes, losing throughput against elastic cross-traffic, e.g., by 54\% than its fair share (\S\ref{s:big-experiment} and \S\ref{s:rw-wl})). The reason is that Copa uses instantaneous measurements for mode-switching, making it vulnerable to variations in the cross traffic.
\item Copa misclassifies cross traffic when the inelastic traffic rate is high, or when elastic flows have high RTTs ($\S$\ref{s:copa-compare}). 
\end{CompactEnumerate}
Moreover, since \name does not rely on properties of any specific control algorithm (e.g., emptying queues every 5 RTTs), it applies to any combination of TCP-competitive and delay-controlling schemes and can be used as a building block.

BBR~\cite{bbr} estimates the bottleneck bandwidth ($b$) and minimum RTT ($d$). 
It paces traffic at a rate $b$ while capping the number of in-flight packets to $2\times b\times d$.
To estimate the bottleneck, BBR periodically increases its rate over $b$ for about one RTT and then reduces it for the following RTT. 
BBR uses this sending-rate pattern to obtain estimates of $b$; specifically, it tests if the bottleneck rate exceeds the current estimate $b$ in the rate-increase phase. However, BBR doesn't use these pulses to infer the nature of cross traffic.

PCC-Vivace~\cite{vivace} uses an online learning algorithm to adapt its sending rate to maximize a utility function that incorporates the achieved rate, delay, and loss rate. Our experiments ($\S$\ref{s:big-experiment}, $\S$\ref{s:rw-wl}) show that Vivace cannot achieve both low delay with inelastic cross traffic and compete fairly with elastic TCP flows.  Compound TCP~\cite{compound} maintains both a loss-based window and a delay-based window, and transmits data based on the sum of the two windows. Compound does not attempt to switch between two modes, and therefore it incurs high queuing delays due to its loss-based window. 
\newcommand{\switchsec}{Cross-Traffic Estimation}
\section{\switchsec}
\label{s:switching}

We present a simple new method to estimate the total rate of cross traffic at the sender (\S\ref{s:zrate}). Then, we show how to detect whether the cross traffic contains {\em any} ACK-clocked elastic flows, describing the key principles (\S\ref{s:zelasticity}) and a practical method (\S\ref{sec:pracelas}).

Figure~\ref{fig:sysmod} shows our network model and introduces some notation. A sender communicates with a receiver over a single bottleneck link of rate $\mu$. The bottleneck link is shared with cross traffic, consisting of an unknown number of flows, each of which is either elastic or inelastic. $S(t)$ and $R(t)$ denote the time-varying sending and receiving rates, respectively, while $z(t)$ is the total rate of the cross traffic. We assume that the sender knows $\mu$, and can use prior work to estimate it ($\S$\ref{s:implementation}).

\subsection{Estimating the Rate of Cross Traffic}
\label{s:zrate}

\begin{figure}[t]
    \centering
    \includegraphics[width=0.8\columnwidth]{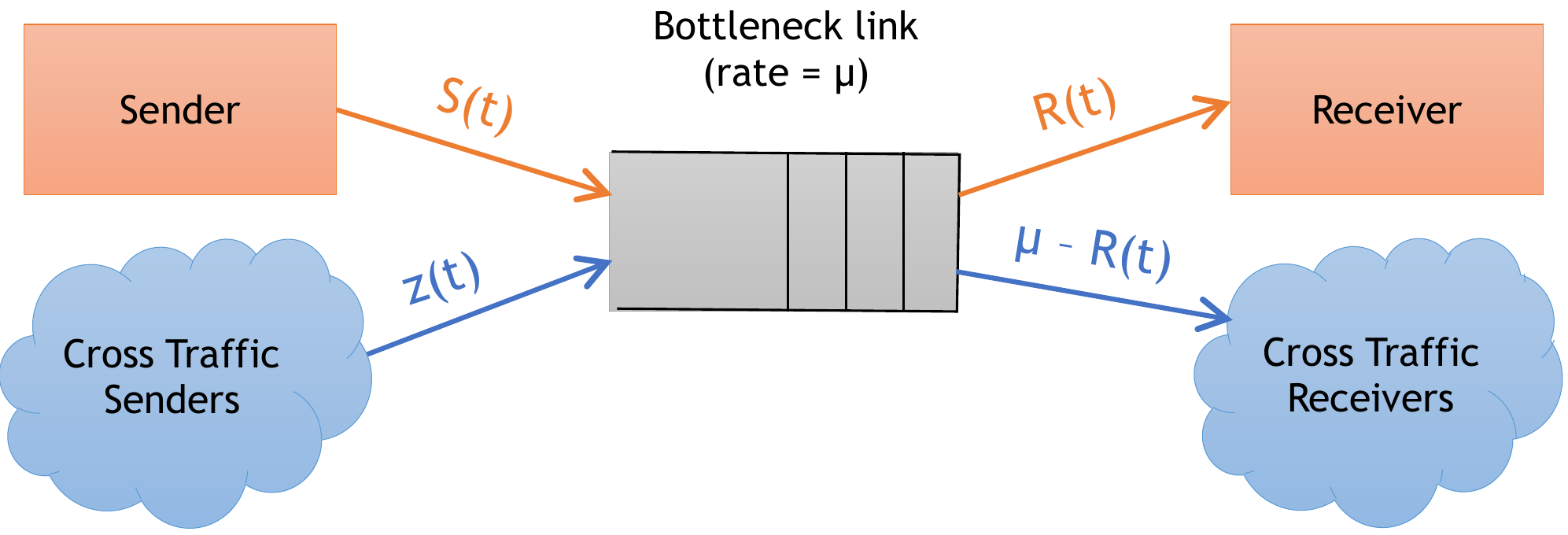}
    \vspace{-2mm}
    \caption{\small{{\bf Network model.} 
    The time-varying total rate of cross traffic is $z(t)$. The bottleneck link rate is $\mu$. The sender's transmission rate is $S(t)$, and the rate of traffic received by the receiver is $R(t)$.}}
    \label{fig:sysmod}
    \vspace{-4.5mm}
\end{figure}

In \Fig{sysmod}, the total traffic into the bottleneck queue is $S(t) + z(t)$, of which the receiver sees $R(t)$. As long as the bottleneck link is busy (\ie its queue is not empty), and the router treats all traffic the same way, the ratio of $R(t)$ to $\mu$ must be equal to the ratio of $S(t)$ and the total incoming traffic, $S(t)+z(t)$. Using this property, we propose a new estimator for $z(t)$:
\begin{align} \label{eq:zfromR}
\hat{z}(t) &= \mu \frac{S(t)}{R(t)} - S(t).
\end{align}

We estimate $S(t)$ and $R(t)$ by considering $n$ packets at a time:

\begin{equation} 
\centering
S_{i,i+n} = \frac{n_{bytes}} {s_{i+n} - s_i}, \qquad R_{i,i+n} = \frac{n_{bytes}} {r_{i+n} - r_i},
\label{eq:SR}
\end{equation}

where $n_{bytes}$ is the number of bytes in the $n$ packets, $s_k$ is the time at which the sender sends packet $k$, $r_k$ is the time at which the sender receives the ACK for packet $k$, and the units of the rates are bytes per second. Note that $S(t)$ and $R(t)$ must be measured over the {\em same} $n$ packets.

We have conducted several tests with various patterns of cross traffic to evaluate the effectiveness of this $z(t)$ estimator. The overall error is small: the 50th and 95th percentiles of the relative error are 1.3\% and 7.5\%, respectively.
Unlike prior work on estimating cross-traffic rate~\cite{strauss2003measurement,jain2002pathload,hu2003evaluation}, our method is in-band and does not use any probe packets; it relies on the property that the sender is persistently backlogged.

\subsection{Elasticity Detection: Principles}
\label{s:zelasticity}

We now turn to designing an online estimator for a sender to determine if the cross traffic includes {\em any} elastic flows.\footnote{Receiver participation will improve accuracy by avoiding the need to estimate $R(t)$ from ACKs at the sender, but would be a little harder to deploy.} A strawman approach might attempt to detect elastic flows by estimating the contribution of the cross traffic to queueing delay. For example, the sender can estimate its own contribution to the queueing delay---\ie the ``self-inflicted'' delay---and if the total delay is significantly higher than the self-inflicted delay, conclude that the cross traffic is elastic. 

 \begin{figure}[t]
     \centering
    \includegraphics[width=1.0\columnwidth]{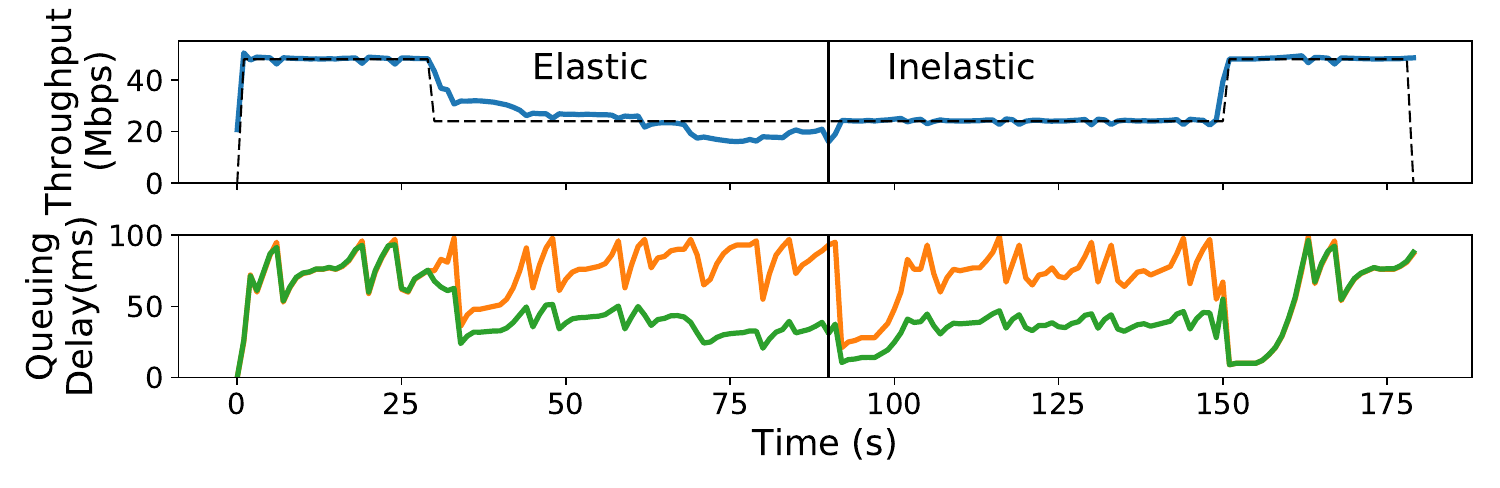}
    \vspace{-9mm}
     \caption{\small Instantaneous delay measurements do not reveal elasticity. The bottom plot shows the total queueing delay (orange) and the self-inflicted delay (green). The experiment contains one background Cubic flow in the elastic region (30--90 s) and CBR cross traffic in the inelastic region (90--150 s).}
     \label{fig:si-delay}
 \end{figure}


This scheme does not work. To see why, consider the experiment in Figure~\ref{fig:si-delay}, where a Cubic flow shares a link with elastic and inelastic  traffic in two separate time periods. The self-inflicted queueing delay for the Cubic flow (green, bottom figure) looks the same in the elastic and inelastic phases. The reason is that a flow's share of the queue occupancy is proportional to its throughput, which is roughly the same in the two phases (top figure). Because the Cubic flow gets 50\% of the bottleneck link, its self-inflicted delay is roughly half of the total queueing delay always (orange, bottom figure). This example suggests that instantaneous measurements cannot be used to distinguish between elastic and inelastic cross traffic.

\smallskip
\noindent{\bf To detect elasticity, tickle the cross traffic!} 
Our method detects elasticity by monitoring how the cross traffic responds to induced traffic variations at the bottleneck link over a period of time. The key observation is that elastic flows react in a predictable way to rate fluctuations at the bottleneck. Consider, for example, long-running Cubic or Reno flows, which are ACK-clocked. For these flows, if an ACK is delayed by a time duration $\delta$, then the next packet transmission will also be delayed by $\delta$. Therefore changes in the rate of packet arrivals at the receiver cause similar changes in the sending rate after one RTT via the ACKs. By contrast, the sending rate of inelastic flows does not depend on the receive rate.

We induce changes in the inter-packet spacing of cross traffic at the bottleneck link by sending packets in {\em pulses}. We take the desired sending rate, $S(t)$, and alternate between sending at rates higher and rates lower than $S(t)$, ensuring that the mean rate is $S(t)$. Sending in such pulses (e.g., modulated on a sinusoid) changes the inter-packet spacing of the cross traffic departing the bottleneck link in a controlled manner. If the cross traffic contains elastic flows, then because of the induced changes in the ACK clocks of those flows, their rates will react to our pulses. When we increase our rate, the elastic cross traffic will reduce its rate in the next RTT, and conversely. If enough of the cross traffic is elastic, then our sender can measure and detect these fluctuations in the cross traffic rate.

\begin{figure}
    \centering
    \begin{subfigure}[b]{0.45\columnwidth}
        \includegraphics[width=\columnwidth]{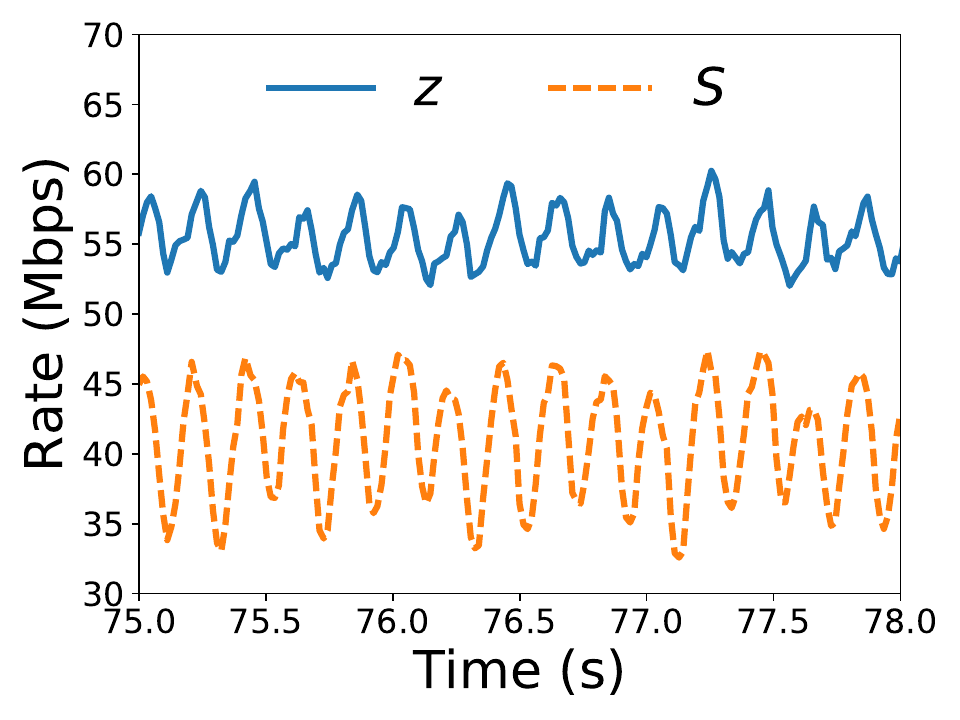}
        \vspace{-6.5mm}
        \caption{\small Elastic cross traffic}
        \label{fig:ts:elastic}
    \end{subfigure}
    \begin{subfigure}[b]{0.45\columnwidth}
        \includegraphics[width=\columnwidth]{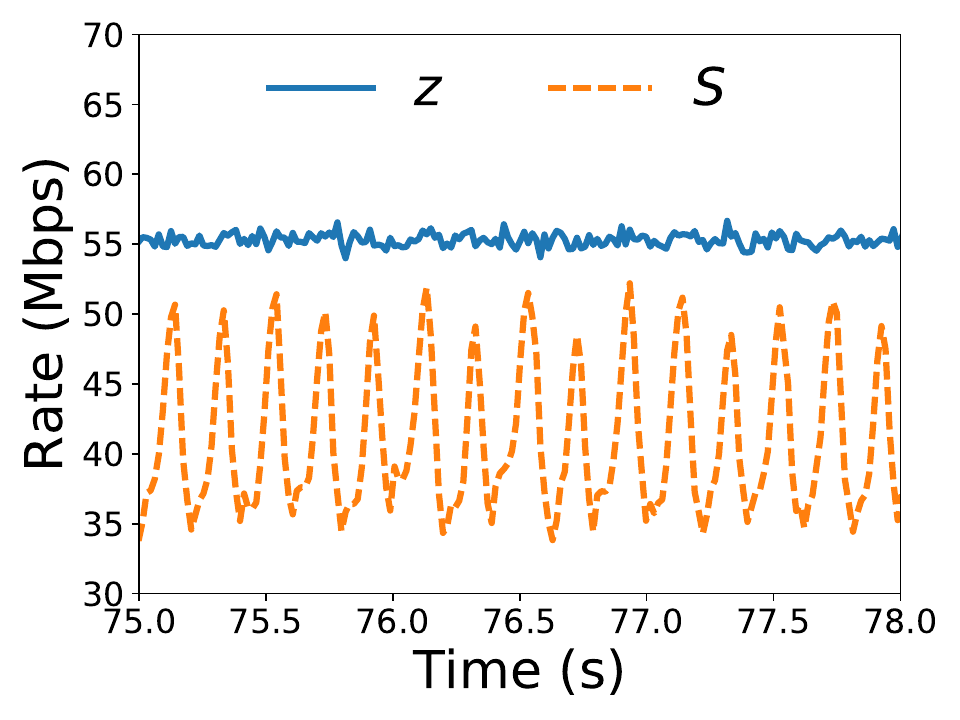}
        \vspace{-6.5mm}
        \caption{\small Inelastic cross traffic}
        \label{fig:ts:inelastic}
    \end{subfigure}
    \vspace{-3.5mm}
    \caption{\small{{\bf Cross traffic's reaction to pulses.} The pulses change the inter packet spacing for cross traffic. Elastic traffic reacts to these changes after a RTT. Inelastic cross traffic is agnostic to these changes.} }
    \label{fig:ts}
    \vspace{-5mm}
\end{figure}

\Fig{ts:elastic} and \Fig{ts:inelastic} compare the responses of elastic (Cubic) and inelastic (constant bit rate) cross traffic when the sender transmits packets in sinusoisal pulses at frequency $f_p=5$~Hz. $S(t)$ is the sender's rate and $z(t)$ is the estimated cross traffic rate computed using Eq.~\eqref{eq:zfromR}. The path has a minimum RTT of 50 ms and a buffer size of 100 ms (2$\times$ the bandwidth-delay product). The elastic flow's sending rate after one RTT is inversely correlated with the pulses in the sending rate, while the inelastic flow's sending rate is unaffected.

\subsection{Elasticity Detection: Practice}
\label{sec:pracelas}
\label{s:pracelas}

To produce a practical method to detect cross traffic using this idea, we must address three challenges:
\begin{CompactEnumerate}
\item Pulses in the sending rate must induce a measurable change in $z$, but not congest the bottleneck link.
\item Because there is natural variation in cross traffic, and noise in $\hat z$, it is not easy to perform a robust comparison between the predicted change in $z$ and the measured $z$. 
\item Because the sender does not know the RTTs of cross-traffic flows, it does not know when to look for the predicted response in the cross-traffic rate.
\end{CompactEnumerate}

The first method we developed to solve these problems measured the {\em cross-correlation} between $S(t)$ and $z(t)$.  A cross-correlation near zero would be considered inelastic cross traffic, whereas a significant non-zero value would indicate elastic cross traffic. 
We found that this approach works well (with square-wave pulses) if the cross traffic is substantially elastic and has a similar RTT to the flow trying to detect elasticity, but not otherwise. The trouble is that because elastic cross traffic will react after {\em its} RTT, $S(t)$ and $z(t)$ must be aligned using the cross traffic's RTT, which is not easy to infer. Moreover, the elastic flows in the cross traffic may have different RTTs, making the alignment even more challenging. 

%
\noindent {\bf From time to frequency domain.} We have developed a method, \name{}, that overcomes the three challenges stated above. It uses two ideas. First, the sender modulates its packet transmissions using {\em sinusoidal pulses} at a known frequency $f_p$, with amplitude equal to a modest fraction (e.g., 25\%) of the bottleneck link rate. 
These pulses induce a noticeable change in inter-packet times at the link without causing congestion, because the queues created in one part of the pulse are drained in the subsequent part, and the period of the pulses is short (e.g., $f_p = 5$~Hz). By using short pulses, we ensure that the total burst of data sent in a pulse is a small fraction of the typical bottleneck queue size. 

Second, the sender looks for periodicity in the cross traffic rate at frequency $f_p$, using a frequency domain representation of the cross-traffic rates. We use the Fast Fourier Transform (FFT) of the time series of the cross traffic estimate $\hat z(t)$ over a short time interval (\eg 5 seconds). Detecting periodicity in the frequency domain is more robust than the time-domain, for the same reason that frequency modulation provides better signal-to-noise ratio than amplitude modulation~\cite{rappaport1996wireless}: it is less affected by variations in the cross traffic rate and measurement noise. 
Further, observing the cross traffic's response at a known frequency, $f_p$, yields a method that is robust to the presence of multiple ACK-clocked flows with different RTTs. All the elastic flows in the cross traffic, irrespective of their RTTs and congestion control protocol, will exhibit rate oscillations at the frequency $f_p$. As a result, there will be an overall response at frequency $f_p$ in the cross traffic, equal to superposition of the responses of the individual elastic flows at frequency $f_p$.\footnote{In theory, the response of flows with different RTTs may cancel each other out, but this is very unlikely since it requires specific combinations of RTTs. We have not seen this problem occur in our experiments (\S\ref{s:robustness-synthetic}).}

\begin{figure}
    \centering
    \includegraphics[width=\columnwidth]{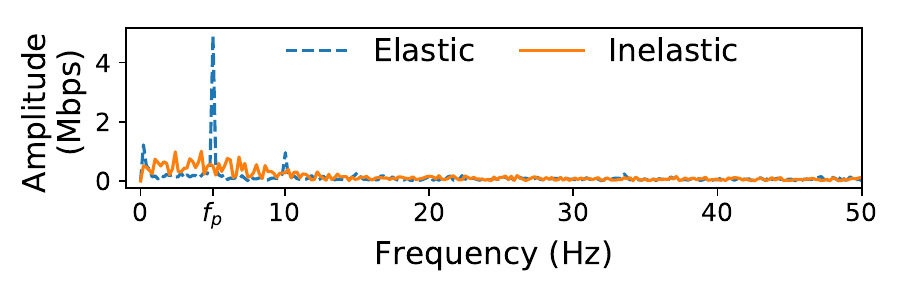}
    \vspace{-10mm}
    \caption{\small{{\bf Cross traffic FFT for elastic and inelastic traffic.} Only the FFT for elastic traffic has a pronounced peak at $f_p$ (5 Hz).}}
    \label{fig:fft}
    \vspace{-6mm}
\end{figure}

\Fig{fft} shows the FFT of the $\hat z(t)$ time-series produced using Eq.~(\ref{eq:zfromR}) for examples of elastic and inelastic cross traffic, respectively. Elastic cross traffic exhibits a pronounced peak at $f_p$ compared to the neighboring frequencies, while for inelastic traffic the FFT magnitude is spread across many frequencies. The magnitude of the peak depends on how much of the cross traffic is elastic; the more elastic the cross traffic, the sharper the peak at $f_p$. Therefore, rather than compare the peak at $f_p$ to a pre-determined threshold, we compare it to the magnitude of the nearby frequencies. 

We define the {\em elasticity metric}, $\eta$, as follows:
\begin{equation} 
\centering
\eta = \frac{|FFT_{z}(f_{p})|}{\max_{f \in (f_p,2f_{p})} |FFT_{z}(f)|}
\label{eq:elasticity}
\end{equation}

Eq.~\eqref{eq:elasticity} compares the magnitude of the FFT at frequency $f_{p}$ to the peak magnitude in the range from just above $f_{p}$ to just below $2f_{p}$. 
If $\eta$ is less than a threshold $\eta_{thresh} (\geq 1$), then the cross traffic is deemed inelastic; otherwise, it is elastic.

\subsection{Setting Parameters for Elasticity Detection}\label{s:elasticity:parameter_tuning}

\begin{figure}[t]
    \centering
    \includegraphics[width=\columnwidth]{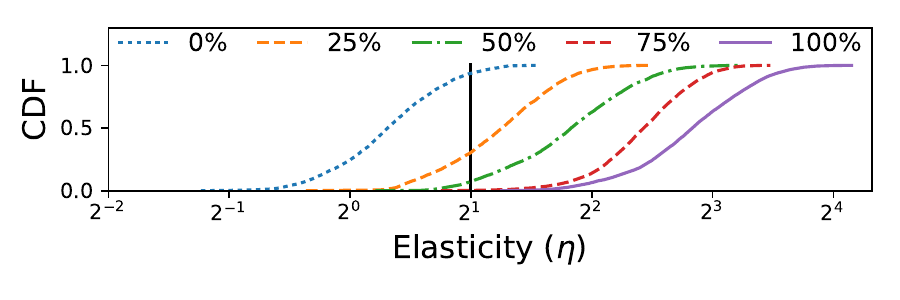}
    \vspace{-10mm}
    \caption{ \small{Distribution of elasticity with varying elastic fraction of cross traffic. The cross traffic consists of an elastic Cubic flow and inelastic Poisson-distributed traffic with different  rates. Completely inelastic cross traffic has $\eta$ close to zero, while completely elastic cross traffic exhibits $\eta$. Cross traffic with some elastic fraction also exhibits high elasticity ($\eta>2$).}}
    \label{fig:elasticity:cdf}
    \vspace{-2mm}
\end{figure}

\noindent{\bf Detection threshold.} In practice, cross traffic is likely to be a mix of elastic and inelastic flows. In such scenarios, we want our detector\cut{\pg{, \name{},}} to be sensitive to the presence of any elastic flows, since even one elastic flow can eventually grab all the link bandwidth from a delay-controlling flow. A large value of $\eta_{thresh}$ will ensure that purely inelastic traffic will always be classified correctly, but cross traffic with small elastic components will be misclassified. \Fig{elasticity:cdf} shows the CDF of elasticity ($\eta$) as the fraction of bytes belonging to elastic flows in the cross traffic varies. The median values range from $\eta = 1$ for purely inelastic traffic to $ \eta = 10$ for purely elastic traffic. We choose a fixed threshold $\eta_{thresh} = 2$, which corresponds to classifying 25\% elastic cross traffic correctly 75\% of the time.

\smallskip
\noindent{\bf FFT duration.} Computing FFTs over a small duration allows quick responses to changes in cross traffic, but it increases errors due to noise. Variations in the inelastic cross traffic over small periods can cause false peaks at $f_p$ in the FFT, causing that traffic to be incorrectly classified. We choose an FFT duration of 5 seconds to balance these concerns.

\begin{figure}[t]
    \centering
    \includegraphics[width=0.8\columnwidth]{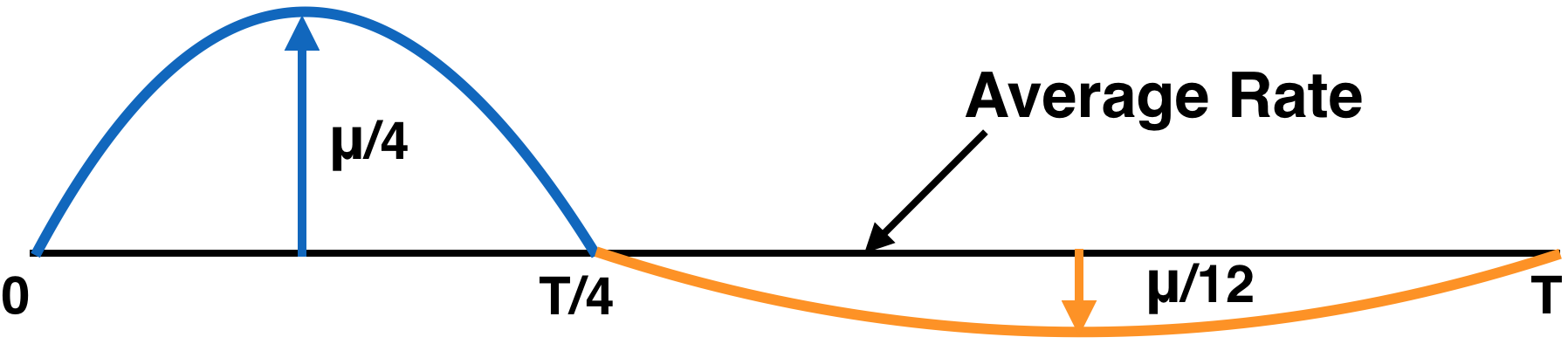}
    \vspace{-3mm}
    \caption{ \small{{\bf Asymmetric sinusoidal pulse.} The pulse has period $T=1/f_p$. The positive half-sine lasts for $T/4$ with amplitude $\mu/4$, and the negative half-sine lasts for the remaining duration, with amplitude $\mu/12$. The two half-sines cancel out each other over one period.}}
    \label{fig:semi-sine}
    \vspace{-6mm}
\end{figure}

\smallskip
\noindent {\bf Pulse shaping.} Rather than a pure sinusoid, we use an {\em asymmetric} sinusoidal pulse, as shown in \Fig{semi-sine}. In the first one-quarter of the pulse cycle, the sender adds a half-sine of a certain amplitude (e.g., $\mu/4$) to $S(t)$; in the remaining three-quarters of the cycle, it subtracts a half-sine with one-third of the amplitude used in the first quarter of the cycle (e.g., $\mu/12$). The reason for this asymmetric pulse is that it enables senders with low sending rates, $S(t)$, to generate pulses. For example, for a peak amplitude of $\mu/4$, a sender with $S(t)$ as low as $\mu/12$ can generate the asymmetric pulse shown in \Fig{semi-sine}; a symmetric pulse with the same peak rate would require $S(t) > \mu/4$. 

Our pulses produce an observable pattern in the FFT when the cross traffic is elastic. Using asymmetric sinusoidal pulses creates harmonics at multiples of the pulse frequency $f_p$. However, these harmonics do not affect $\eta$ (see Eq.~\eqref{eq:elasticity}), which only uses the FFT in the frequency band $[f_p, 2f_p)$.

\smallskip
\noindent {\bf Pulse duration.}
What should the duration, $T$, of the pulse be? The answer depends on two factors: first, the interval over which $S$ and $R$ are measured (with which the sender computes $\hat z$), and second, the amount of data we are able to send in excess of the mean rate without causing congestion. If $T$ were smaller than the measurement interval of $S$ and $R$, the perturbation to the cross traffic rate during one part of the pulse will be averaged out during the rest of the pulse, resulting in no impact on $\hat z(t)$.
But $T$ cannot be too large because the sender transmits in excess of the mean rate $S(t)$ for $T/4$. 
In particular, the size of the burst sent in a pulse is $\frac{2}{\pi}\frac{\mu}{4}\frac{T}{4} = \frac{T\mu}{8\pi} \approx 0.04 \mu T $. If $T$ is equal to the RTT, this is 4\% of the bandwidth-delay product (BDP).

We set $T$ to a large RTT value observed on the Internet, for example $T = 200$ ms, with the rationale that router buffers are typically provisioned to avoid packet losses for one such RTT, and because our implementation measures $S$ and $R$ over one RTT. We measure rates over one RTT because sub-RTT measurements are confounded by burstiness in packet transmissions (e.g., caused by ACK compression~\cite{jiang2003source}). 

If the cross traffic reacts slower than the pulse duration, \name{} might misclassify those flows. A longer pulse duration, corresponding to the response timescale of the elastic traffic, could detect such flows. 
However, longer pulses would also send more traffic into the network and might cause congestion. 
We evaluate this alternative for detecting PCC-Vivace, a rate-based scheme (not ACK-clocked), in \App{slow-react}.

\newcommand{\nimbusprotcol}{Nimbus}
\section{NimbusCC}
\label{s:nimbus-protocol}

\name{}CC is a congestion control system that uses mode switching. It has a TCP-competitive mode in which the sender transmits using a TCP-competitive congestion control algorithm (e.g., Cubic), and a delay-control mode that uses a \dcc algorithm (e.g., Copa). \name{}CC switches between the two modes using our elasticity detector, \name{}.

\subsection{Mode Switching}
\label{s:mode_switching}
At any given time, \name{}CC transmits data at the time-varying rate dictated by the congestion control algorithm running at that time. It modulates this rate with asymmetric sinusoidal pulses (\Fig{semi-sine}). \name{}CC uses the pulsing parameters described in $\S$\ref{s:elasticity:parameter_tuning}, calculating $S$ and $R$ over one window's worth of packets. It computes the FFT for the $z$ measurements reported in the last 5 seconds to calculate elasticity ($\eta$) using Eq. \eqref{eq:elasticity}, and it picks the mode by comparing $\eta$ to $\eta_{thresh} = 2$ (\S\ref{s:elasticity:parameter_tuning}).

We support Cubic and NewReno for the TCP-competitive mode and Copa's default mode and Vegas for the delay-control mode. We also implemented a basic delay-controlling algorithm, BasicDelay, using our cross traffic rate estimator.

Let $S$ be the sending rate and $\hat z$ be the estimated cross-traffic rate, both measured over the last window of packets. Also, let $x$ be the current RTT, and $x_{min}$ be the minimum observed RTT. Upon receiving an ACK, BasicDelay sets its current rate to:
\vspace{-1mm}
\begin{equation}
 \text{Rate} \leftarrow
    S 
    + \alpha (\mu - S - \hat z) 
    + \beta\frac{\mu}{x}
        (x_{\min} + d_t  - x),
\label{eq:delayrule}
\end{equation}
where $\alpha$ and $\beta$ are constants smaller than 1, and $d_t$ is a target queuing delay. The term $(\mu - S - z)$ is the sender's estimate of the spare capacity in the last RTT. By adding an $\alpha$-fraction of the spare capacity to $S(t)$, BasicDelay tries to get closer to the ideal rate. The second term in the above rule seeks to maintain a specified queuing delay, $d_t$, to prevent the queue from both growing too large or going empty. Recall that our cross traffic estimator, Eq.~\eqref{eq:zfromR}, requires a non-empty queue to estimate $z$.

\name{}CC takes special care in initializing the rate when switching to TCP-competitive mode.
\name{}CC sets the rate (and equivalent window) to the rate that was used 5 seconds ago because the elasticity detector takes 5 seconds (FFT Duration) to detect elastic cross traffic. 
During this time, the elastic traffic could cause a reduction in the delay-control mode's rate. 
Hence, \name{}CC resets its rate to the rate at the beginning of the 5-second detection period.

\subsection{Implementation}
\label{s:implementation}

We implemented \name{}CC using CCP~\cite{ccp-sigcomm18}, which provides a convenient way to express the signal processing operations in user-space code.
It uses estimates of $S$, $R$, the RTT, and packet losses from the Linux kernel every 10 ms.

Calculating $\hat z$ requires an estimate of the bottleneck link rate ($\mu$). There has been much prior work~\cite{packettrain, dovrolis2001packet, downey1999using, lai2000measuring, jacobson1997pathchar, lai2001nettimer, mar2000pchar} in estimating $\mu$, which \name{}CC could use.
We use the maximum received rate as the estimate, taking care to avoid incorrect estimates due to ACK compression. We evaluate the impact of errors in estimating $\mu$ on elasticity detection in \Sec{robustness-synthetic}.
\section{Visualizing NimbusCC}
\label{s:switching-illustration}

\label{s:big-experiment}

\begin{table}[tp!]
\small
\begin{center}
\begin{tabular}{cccc}
 Scheme & Throughput $\Delta$ & Throughput $\Delta$ & QDelay\\ & Elastic & Inelastic & Inelastic\\ 
 \hline
  \hline
\name{}CC & $-10\%$ & $0\%$ & $12$\,ms \\
Cubic+BasicDelay & & & \\
  \hline
\name{}CC & $-15\%$ &  $-1\%$ & $14$\,ms \\
Cubic+Copa & & &\\
  \hline
Cubic & $+12\%$ & $0\%$ & {\color{red}$78$\,ms} \\
 \hline
BBR & {\color{red}$+61\%$} & $-2\%$ & {\color{red}$56$\,ms}\\
  \hline
Vegas & {\color{red}$-79\%$} & {\color{orange}$-15\%$}  & $3$\,ms\\
  \hline
Compound &{\color{orange}$-25\%$} & $0\%$ & {\color{orange}$45$\,ms}\\
  \hline
Copa & {\color{red}$-54\%$} & {\color{orange}$-19\%$} & $18$\,ms\\
  \hline
PCC-Vivace & {\color{red}$+61\%$} & $-2\%$ &  {\color{orange}$27$\,ms}\\
\end{tabular}
\end{center}
\caption{\small{Average queuing delay (in ms) in the inelastic region, and deviation from fairshare throughput in elastic and inelastic regions from \Fig{big}. \name{}CC is the only scheme to achieve close to fair-share throughput and low delays.}}
\label{tab:big-exp}
\vspace{-9mm}
\end{table}

\begin{figure}[t]
    \centering
		\includegraphics[width=\linewidth]{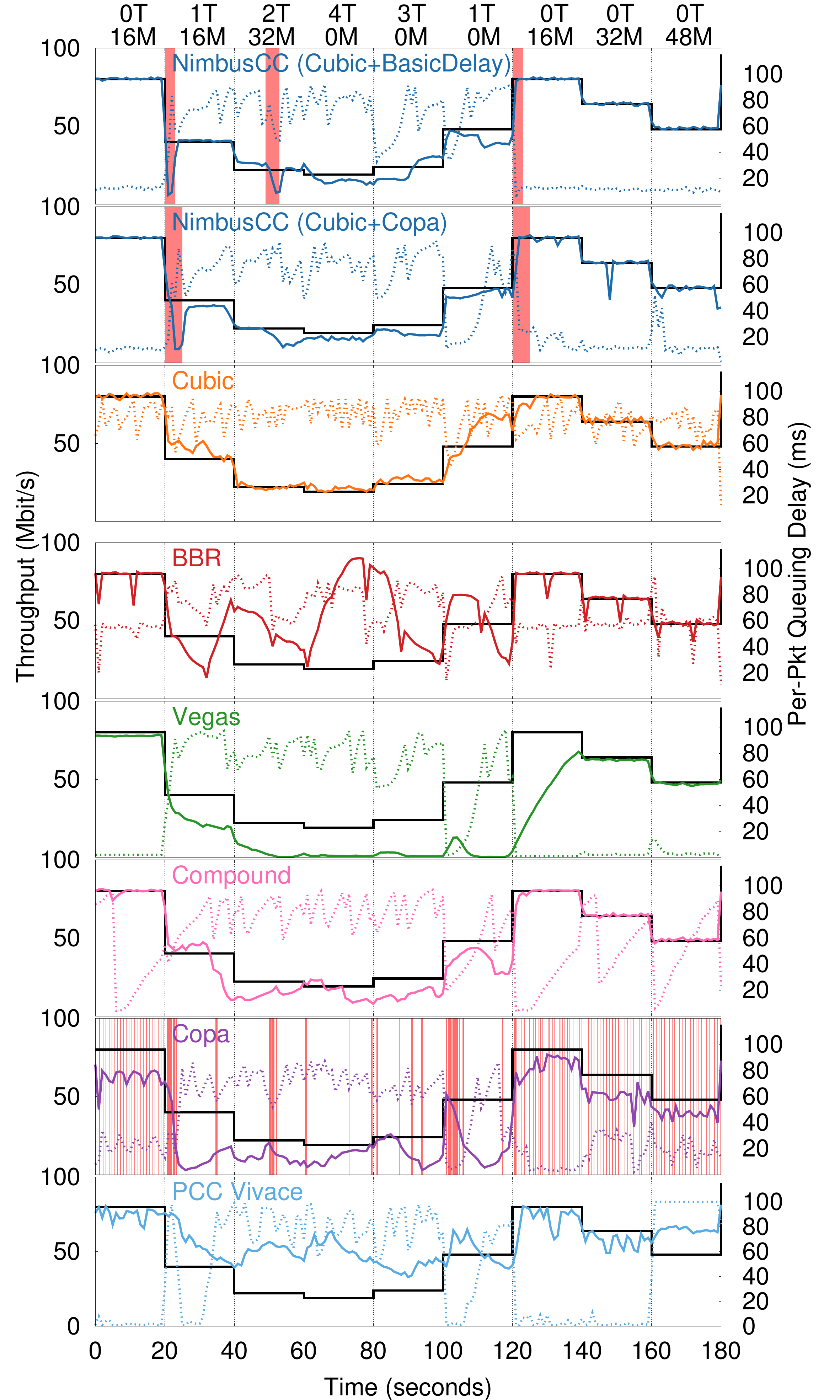}
		\vspace{-7mm}
        \caption{\small Performance on a 96 Mbit/s \mahimahi link with 50 ms delay and 2 BDP of buffering while varying the rate and type of cross traffic as denoted at the top of the graph.
        $x$M denotes $x$ Mbit/s of inelastic Poisson cross-traffic. $y$T denotes $y$ long-running Cubic cross-flows.
        The solid black line indicates the correct time-varying fair-share rate that the protocol should achieve given the cross-traffic. For each scheme, the solid line shows throughput and the dotted line shows queuing delay. The cross-traffic contains elastic flows from 20--120 s. For \name and Copa, the red shaded regions indicate times spent in the wrong mode (\eg \dcc with elastic cross traffic).}
        \vspace{-6mm}
    \label{fig:big}
\end{figure}

We illustrate \name{}CC on a synthetic workload with time-varying cross traffic. We emulate a bottleneck link in \mahimahi~\cite{mahimahi}, a link emulator. The network has a bottleneck rate of 96 Mbit/s, a minimum RTT of 50 ms, and 100 ms (2 BDP) of buffering. We compare two mode-switching protocols, \name{}CC (Cubic+BasicDelay) and \name{}CC (Cubic+Copa), with Cubic, BBR, Vegas, and PCC-Vivace (all from Linux), Copa (from Copa's authors), and Compound atop CCP (written by us).

The cross traffic varies over time between elastic, inelastic, and a mix of the two. 
We generate inelastic cross-traffic using Poisson packet arrivals at the specified mean rate.
Elastic cross-traffic uses Cubic, via {\ct iperf}~\cite{iperf}.

\Fig{big} shows the throughput and queuing delays for the various protocols, as well as the correct fair-share rate. Table ~\ref{tab:big-exp} summarizes the deviation from fair-share throughput in the elastic (20--120 s) and inelastic (0--20 and 120--180 s) regions, and the mean queuing delay in the inelastic region. The delay in the elastic region is similar for all schemes.

Throughout the experiment, both \name{}CC variants achieve throughput close to the fair-share rate and low ($\leq$15 ms) queuing delays in the presence of inelastic cross traffic. 
With elastic cross traffic, both variants switch to TCP-competitive mode within 5 seconds and achieve close to their fair share. The delays during this period approach the buffer size because the competing traffic is buffer-filling; the delays return to their previous low value (15 ms) within 5 seconds after the elastic flows complete. \name{}CC stays in the correct mode throughout the experiment, except for one interval in the elastic period. The deviation from fair-share in the elastic region is because Cubic is not perfectly fair to itself over short time periods.


Cubic achieves it's fair-share rate but experiences high delays (80 ms) throughout. BBR's throughput is often much higher than its fair share with high delays even against inelastic cross-traffic, which prior work has also observed~\cite{copa,bbr-evaluation}.

Vegas suffers from low throughput in the presence of elastic cross-traffic as it reacts to packet delays. Compound ramps up its rate quickly when it detects low delays, but behaves like TCP Reno otherwise. Hence, it attains slightly lower than its fair-share rate in the presence of Cubic flows, and suffers from high delays even with inelastic cross-traffic. 

Unlike Nimbus, Copa uses instantaneous measurements for mode-switching, and is vulnerable to the variations in cross traffic. As a result, while Copa generally uses the correct mode it frequently switches mode unnecessarily; Copa makes 28 switching errors in the elastic region, while \name{}CC only switches once.
In the elastic period, Copa's frequent mode switches lower its throughput (14 Mbit/s) compared to \name{}CC (27.5 Mbit/s) and fair-share rate (\eg see 100--120 s). 
Further, by draining queues periodically, Copa incurs minor underutilization against inelastic traffic (\eg 140--160 s).

Vivace competes unfairly with elastic traffic. At times, Vivace fails to maintain low delays against inelastic cross traffic and  incurs heavy packet loss (\eg 160--180s). 

\newcommand{\multinimbussec}{Multiple \name{}CC Flows}
\section{\multinimbussec}
\label{s:multinimbus}

What happens when a bottleneck is shared by multiple \name{}CC flows? Ideally, we want all the \name{}CC flows to remain in delay-control mode when there is no elastic cross traffic, and compete well with elastic cross traffic otherwise. 

One approach is for the \name{}CC flows to all pulse at the same frequency. However, in this case, they will all detect a peak in the FFT at the oscillation frequency. They will all then stay in TCP-competitive mode and won't be able to maintain low delays, even when there is no elastic cross traffic. A second approach is for different \name{}CC flows to pulse at different frequencies. But this approach cannot scale to more than a few flows, because the set of distinguishable frequencies is limited (recall that the pulse period $T$ cannot be too small). 

\smallskip
\noindent{\bf The pulser and the watchers.}
We propose a third approach. One of the \name{}CC flows assumes the role of the {\em pulser}, while the others are {\em watchers}. They coordinate with no explicit communication; in fact, each \name{}CC flow is unaware of the identities, or even existence, of the others.

The \pulser sends data by modulating its rate with asymmetric sinusoids. The \pulser uses two different frequencies, $f_{pc}$ in TCP-competitive mode, and $f_{pd}$ in delay-control mode. The values of these frequencies are fixed and agreed upon beforehand; we use $f_{pc}=5$ Hz and $f_{pd}=6$ Hz in our experiments.\footnote{These values are in accordance with bounds on $T$ and $f$ described in \S\ref{s:nimbus-protocol}.} 

A \watcher infers whether the \pulser is pulsing at frequency $f_{pc}$ or frequency $f_{pd}$ by computing the FFT of its receive rate, $R$, at these two frequencies. It then picks the mode corresponding to the larger peak to match the pulser's mode. Note that since a watcher is not pulsing, it can detect the pulser's pulses in its own receive rate, $R$; i.e., it does not even need to estimate $z$. The pulser, on the other hand, cannot look at its own $R$ to detect pulses in the cross traffic, since it will end up detecting its own pulses.

For multiple \name{}CC flows to maintain low delays during times when there is no elastic cross traffic on the link, the \pulser must classify \watcher traffic as inelastic. Note that from the \pulser's perspective, the \watcher flows are part of the cross traffic; thus, to avoid confusing the \pulser, the rate of \watchers must not react to the pulses of the \pulser. To achieve this goal, a watcher applies an exponentially weighted moving average (EWMA) filter to its transmission rate before sending data. The EWMA filter cuts off all frequencies in the sending rate that exceed $\min(f_{pc},f_{pd})$.

\smallskip
\noindent
{\bf Pulser election.}
A distributed and randomized election decides which flow is the \pulser and which are \watchers. If a \name{}CC flow determines that there is no \pulser (by seeing that there is no peak in the FFT at the two potential pulsing frequencies), then it decides to become a pulser with a probability proportional to its transmission rate:
\begin{equation}
   p_{i} = \frac{\kappa\tau}{\textnormal{FFT Duration}} \times \frac{R_{i}}{\mu}.
   \label{eq:pi}
\end{equation}
Each flow makes decisions periodically, \eg every $\tau = 10$ ms, $\kappa$ is a constant, and $R_i$ is the receive rate of the $i^{th}$ flow. This rule ensures that the expected number of flows that become pulsers over the FFT duration is at most $\kappa$. To see why, note that the expected number of pulsers is equal to the sum of the probabilities in Eq.~\eqref{eq:pi} over all the decisions made by all flows in the FFT duration. Since $\sum_i R_i \leq \mu$ and each flow makes ($\textnormal{FFT Duration} / \tau$) decisions, these probabilities sum up to at most $\kappa$.

It is also not difficult to show that the number of pulsers within an FFT duration has  approximately a Poisson distribution with a mean of $\kappa$~\cite{probabilitybook}. Thus the probability that after one flow becomes a pulser, a second flow also becomes a pulser before it can detect the pulses of the first flow in its FFT measurements is $1 - e^{-\kappa}$. Therefore, $\kappa$ involves a tradeoff: a smaller $\kappa$ will lead to fewer conflicts but will take longer to elect a pulser.

For any value of $\kappa$, there is a non-zero probability of more than one concurrent pulser. If there are multiple pulsers, then each pulser will observe that the cross traffic has more variation than the variations it creates with its pulses. This can be detected by comparing the magnitude of the FFT of the cross traffic $z(t)$ at $f_p$ with the FFT of the pulser's receive rate $R(t)$ at $f_p$.  If the cross traffic's FFT has a larger magnitude at $f_p$, the \name{}CC pulser concludes that there must be multiple pulsers and switches to a watcher with a fixed probability.

\newcommand{\limitsec}{Limitations}
\section{\limitsec}
\label{s:limit}
\begin{table}
\small
\begin{center}
\begin{tabular}{cccc}
 Cross Traffic & Elastic & ACK-Clocked & Classification\\ 
 \hline
 \hline
 Cubic & Yes & Yes & Elastic \\
  \hline
 Reno & Yes & Yes & Elastic \\
  \hline
 Copa & Yes & Yes & Elastic \\
  \hline
 Vegas & Yes & Yes & Elastic \\
  \hline
 BBR & Yes & If CWND-limited & Elastic*\\
  \hline
 PCC-Vivace & Yes & No & Inelastic* \\
  \hline
 Fixed window & Yes & Yes & Elastic \\
  \hline
 App. limited & No & No & Inelastic \\
  \hline
 Const. stream & No & No & Inelastic \\
\end{tabular}
\end{center}
\caption{\small{\bf Classification by Nimbus.}}
\label{tab:elasticity}
\vspace{-11mm}
\end{table}

Table~\ref{tab:elasticity} summarizes how \name{} classifies different types of cross traffic.
Recall that our method relies on the cross traffic responding to variations induced by pulses on an RTT timescale. This is true of all ACK-clocked protocols, which are classifed as elastic. 

\name{} does not always classify BBR cross traffic as elastic. When the buffer is large, BBR becomes ACK-clocked and is elastic. 
However, when the buffer is small, BBR responds on timescales longer than an RTT; here, \name{} classifies it as inelastic. 
Nonetheless, we find that \name{}CC (with Cubic as the TCP-competitive protocol) achieves similar throughput to Cubic when competing against BBR (Appendix $\S$\ref{app:bbr-comparison}). 

Rate-based protocols, (\eg PCC-Vivace) may not react on RTT timescales.
For example, \name{} in its default configuration classifies PCC-Vivace as inelastic because it does not react quickly enough to \name{}'s pulses.  
Increasing the pulse duration
helps \name{} to correctly classify such flows as elastic (Appendix~\ref{app:slow-react}). 
Increasing the pulse duration might of course also increase queuing delays. 
Since most elastic traffic today is ACK-clocked, we use a small pulse duration by default. 
In the future, if rate-based protocols become widely deployed, the pulse duration could be adjusted accordingly. 

The detector also assumes that the flow has a single bottleneck. Multiple bottlenecks can add noise to \name{}'s rate measurements, preventing accurate cross-traffic estimation. The challenge is that the spacing of packets at one bottleneck is not preserved when traversing the second bottleneck.

While \name{} can detect presence of elastic flows, it cannot detect the specific congestion protocol used by competing flows. If the TCP-competitive protocol \name{}CC uses is different from that of the cross traffic, there could be unfairness. Further, if elastic cross traffic is using a \dcc scheme like Vegas, then \name{} could miss out on an opportunity to control delays if it uses a buffer-filling TCP-competitive algorithm. Detecting the congestion control protocols used by competing elastic flows remains an open question. 


\newcommand{\evalsec}{Evaluation}
\section{\evalsec}
\label{s:eval}

We evaluate our elasticity detection method, \name{}, and a specific protocol using \name{}CC: Cubic+BasicDelay, as in \Sec{mode_switching}. We use the Mahimahi emulator and measure the performance benefits (\Sec{rw-wl}), robustness (\Sec{copa-compare}), and fairness (\Sec{multiple-nimbus-fair-sharing}) of elasticity detection with realistic traffic workloads. We also evaluate the performance of \name{}CC on real Internet paths (\Sec{realworld}). All experiments use our Linux implementation (\Sec{implementation}). 

\subsection{\name{}CC Benefits from Elasticity Detection}
\label{s:other-bundle-transports}
\label{s:rw-wl} 

\begin{figure}[t]
    \centering
\begin{subfigure}[t]{\columnwidth}
    \centering
    \includegraphics[width=\columnwidth]{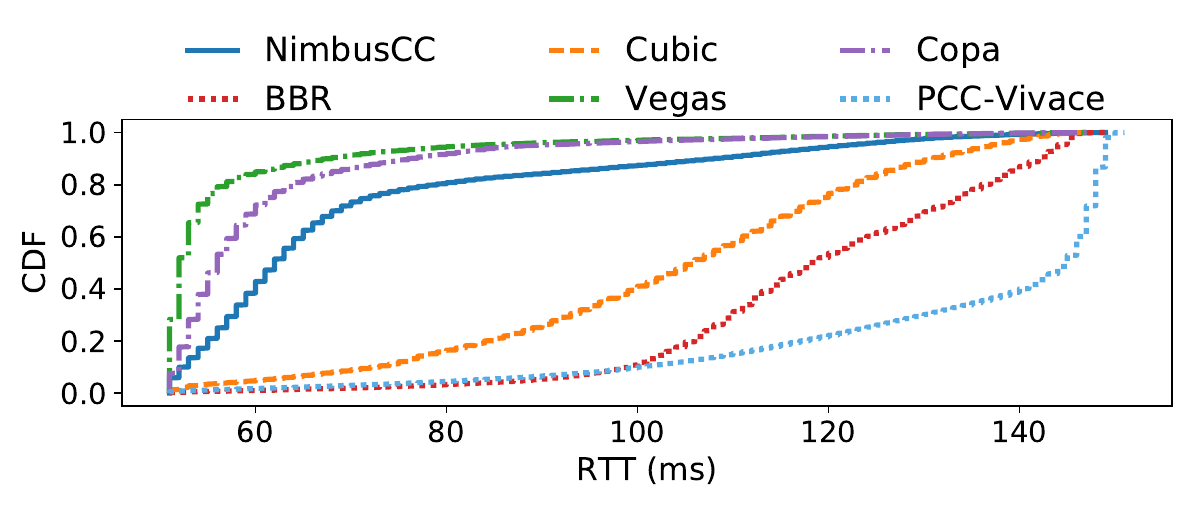}
    \vspace{-5mm}
    \caption{\name{}CC reduces delay relative to Cubic, BBR and PCC-Vivace. It has higher delays than Copa and Vegas, but those two schemes have lower throughput than the fair share against elastic cross-traffic (see figure below).}
    \label{fig:emp:rtt}
\end{subfigure}

\begin{subfigure}[t]{\columnwidth}
    \centering
    \includegraphics[width=\columnwidth]{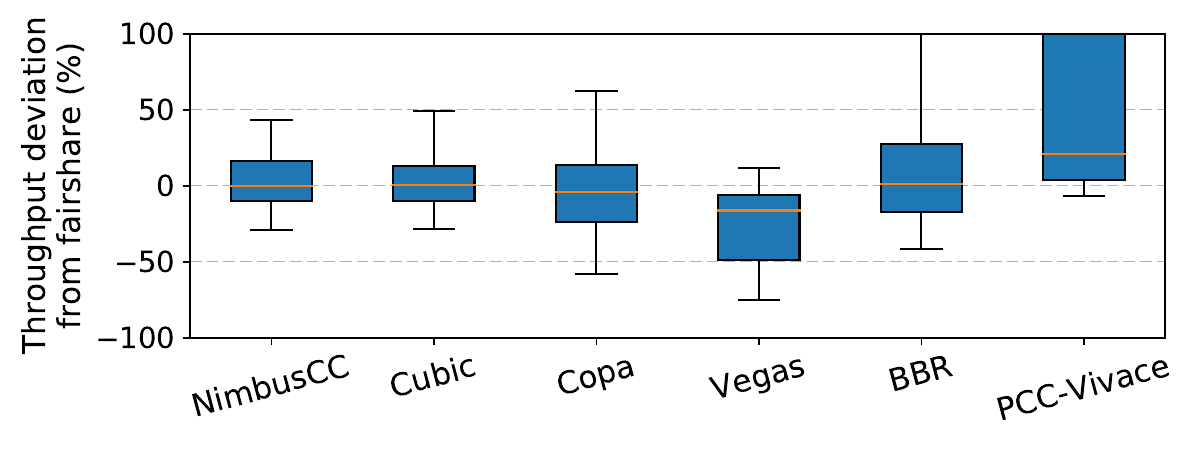}
    \vspace{-5mm}
    \caption{\small{Deviation in throughput from fair-share. \name{}CC and Cubic achieve highest fairness. Vegas and Copa deviate from fair-share at lower percentiles and lose throughput; BBR and PCC-Vivace are significantly higher than fair share. Midline is the median, the box edges are the 25\%ile and 75\%ile, the whisker notches are 10\%ile and 90\%ile.}}
    \label{fig:emp:box}
\end{subfigure}

\begin{subfigure}[t]{\columnwidth}
    \centering
    \includegraphics[width=\columnwidth]{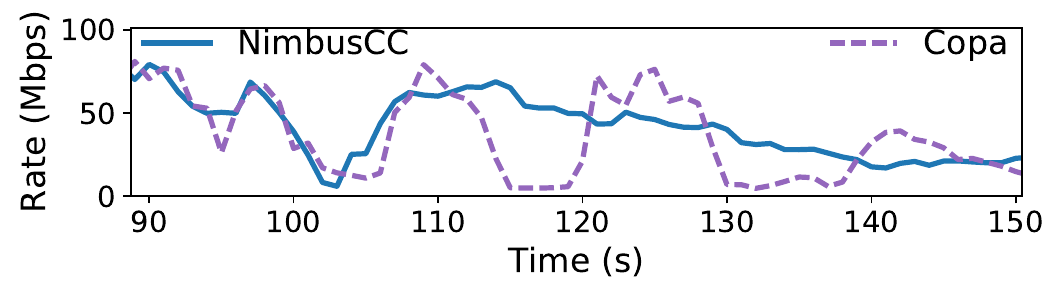}
    \vspace{-5mm}
    \caption{\small{Copa incorrectly switches to its default delay-control mode even when competing against elastic traffic, unlike \name{}CC.}}
    \label{fig:emp:ts}
    \vspace{-3mm}
\end{subfigure}
\caption{Performance of \name{}CC on a cross-traffic workload derived from a packet trace collected at a WAN router.}
\vspace{-5mm}
\end{figure}

We evaluate the delay and throughput benefits of mode switching using trace-driven emulation. We generate cross traffic from an empirical distribution of flow sizes derived from a wide-area packet trace from CAIDA~\cite{caida-dataset}.
This packet trace was collected at an Internet backbone router on January 21, 2016 and contains over 30 million packets recorded over 60 seconds. 
We generate Cubic cross-flows with flow sizes drawn from this data, with flow arrival times generated by a Poisson process to offer a fixed average load to fill 50\% of the link (48 Mbit/s).
Since the flow size distribution is heavy-tailed, the traffic trace consists of periods with {\em a mix of elastic and inelastic cross-traffic,} along with periods with only inelastic cross-flows.

One backlogged flow running a fixed algorithm (\name{}CC, Cubic, Copa, Vegas, PCC-Vivace or BBR) and the cross-traffic flows share a 96 Mbit/s \mahimahi bottleneck link with a propagation RTT of 50 ms and a buffer size of 100 ms. For BasicDelay we used $\alpha=0.8$, $\beta=0.5$ and $d_t=12.5$ ms.

\smallskip 
\noindent\textbf{\name{}CC reduces delays while achieving fair-share throughput.} 
\Fig{emp:rtt} shows the distribution of per-packet RTT and \Fig{emp:box} shows the deviation from fair-share throughput (over 5-second intervals) for various schemes. 
High deviation, 
\ie unfairness with cross traffic, can harm application performance.
For example, for a video application, temporary throughput drops can cause stalls, hurting user experience. 

\name{}CC and Cubic achieve the lowest deviation from fair share among these schemes. \name{}CC's deviation profile is comparable to Cubic (note that both \name{}CC and Cubic deviate from the fair share since Cubic is not perfectly fair to itself over short time periods).
The reason is that \name{}CC correctly switches to Cubic mode in the presence of elastic flows. Additionally, by switching to delay-controlling mode in the absence of elastic flows, \name{}CC achieves lower RTTs, with a median delay only 10 ms higher than Vegas and $>$50 ms lower than Cubic and BBR.

\noindent\textbf{Cost of incorrect mode-switching.} 
Copa has a slightly lower median delay than \name, but at a high cost:
its throughput deviates significantly from the fair-share at the 10$^{\text{th}}$ and 25$^{\text{th}}$ percentiles.   
\Fig{emp:ts} demonstrates why, comparing \name{}CC and Copa during a 60-second interval. Copa often incorrectly operates in its default delay-controlling mode against elastic cross-traffic (\eg 115--120, 130--140 s). 

%
%
Note that in this experiment we would expect the mean throughput for \dcc schemes to be the fair-share, since the total number of bytes in the cross traffic is fixed.
The cross-traffic flow sizes are fixed, and a flow can last for different time durations depending on its throughput. 
Since the flow sizes are fixed, fair co-existence of \name{}CC with elastic cross-fows increases the lifetimes of those flows relative to Copa.
%
This results in \name{}CC achieving lower (but fairer) throughput than Copa during periods (\eg 120--130 s) when an elastic flow has completed in Copa, but not in \name{}CC. 
Moreover, since elastic flows last longer in \name{}CC, the delay is higher than Copa at the tail.

\noindent\textbf{\name{}CC helps cross traffic.} The $95^{\text{th}}$ percentile flow completion time (FCT) of cross-traffic flows reduces by 3-4$\times$ compared to BBR, and 1.3$\times$ compared to Cubic for short ($\leq$ 15 KB) flows (Appendix~\ref{app:cross-traffic-impact}).
In contrast, PCC-Vivace is unfair to the background flows (positive deviation from fairshare). It grabs significantly more bandwidth than all the other schemes and keeps the buffer near-full more than half the time. 
The result is that many background flows do not complete, and their completion times are over 100$\times$ worse than with other schemes. PCC-Vivace also shows higher delays that any other scheme; the median delay is $90$ ms higher than \name{}CC.

We repeated the above experiment with cross-traffic BBR flows instead of Cubic. 
Again, \name{}CC achieves a throughput profile similar to Cubic while reducing delays (Appendix~\ref{app:bbr-comparison}).

\smallskip
\label{s:elasticity-in-real-cross-traffic}
\begin{figure}[t]
    \centering
    \includegraphics[width=\columnwidth]{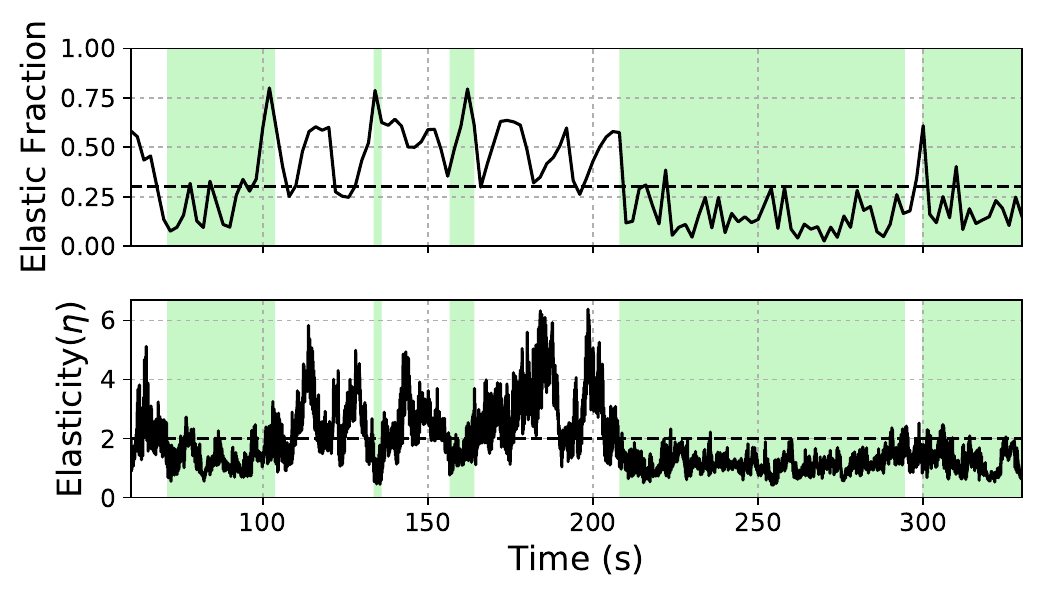}
    \vspace{-5mm}
    \caption{\small{The elasticity metric closely tracks elastic cross-traffic (ground truth measured independently from the rate of ACK-clocked flows). Green-shaded regions indicate inelastic periods.}}
    \label{fig:emp:switch}
    \vspace{-1mm}
\end{figure}

\noindent\textbf{Elasticity detection is accurate compared to ground truth.}
To define ground truth, we note that
short flows (< 10 packets) transmit all data at once, without any rate adjustments.
We thus classify a cross-flow as elastic if it is larger than the initial congestion window of 10 packets, finishing in greater than a \RTT.

The top chart in \Fig{emp:switch} shows the fraction of bytes belonging to elastic flows as a function of time. 
The bottom chart shows the output of the elasticity detector with the dashed threshold line at $\eta=2$. 
The shading corresponds to periods when \name{}CC is in delay-control mode. 
Shaded regions correlate well with the periods when the true fraction of elastic traffic is low (\eg $<0.3$), while white regions correlate well with periods when the elastic fraction is high. 
Unlike Copa, our elasticity detector observes fluctuations in cross traffic \emph{over a period of time in the frequency domain}, and the accuracy is less susceptible to variations in the cross traffic rate. Despite the churn in cross-traffic flows, the overall accuracy of our elasticity detector is over 90\%.

\label{s:videocross}

\begin{figure}
    \centering
    \includegraphics[width=\columnwidth]{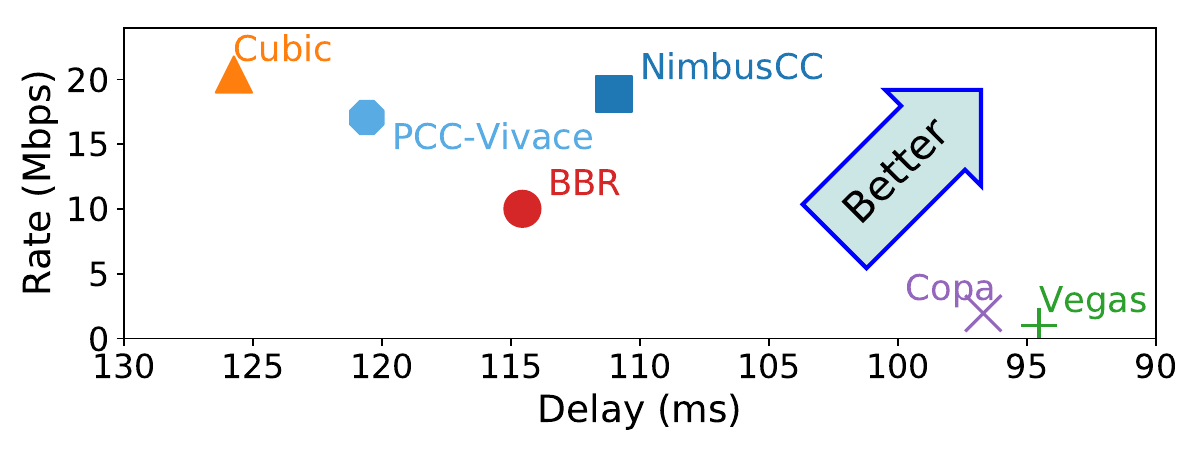}
    \vspace{-4.3mm}
    \caption{\small{Throughput and mean delay (lower delay on the right) with video cross traffic. \name{}CC achieves similar throughput as Cubic but reduces delays and performs better than the other schemes. Copa and Vegas achieve low throughput.}}
    \label{fig:vc}
\end{figure}

\smallskip
\noindent\textbf{Performance with video cross traffic.}
Video streams, 
a large fraction of Internet traffic~\cite{ciscovideo},
can be application-limited (inelastic) or network-limited (elastic) at different points in time. 
We compare the performance of congestion-control algorithms running against cross traffic consisting of a 4k DASH~\cite{dash} video stream using Cubic on a 48 Mbit/s link with 50 ms propagation RTT.
\Fig{vc} shows the throughout and delay of the various schemes. 
Because of effective mode switching, \name{}CC achieves similar throughput as Cubic at 15 ms lower delay. \name{} recognizes application-limited video traffic as inelastic, allowing the sender to control delays in those cases; it rarely recognizes network-limited elastic traffic as inelastic, so does not wrongly reduce its rate as Copa does.

\subsection{Robustness of Elasticity Detection}
\label{s:robustness-synthetic}
We evaluate the robustness of \name{} under a variety of network and traffic conditions.
Unless specified otherwise, we run \name{}CC as a backlogged flow on a 96 Mbit/s bottleneck link with a 50 ms propagation RTT and a 100 ms drop-tail buffer (2 BDP).
%
We consider three categories of synthetic cross-traffic sharing the link with \name{}CC: (i) inelastic Poisson-distributed traffic; (ii) fully elastic traffic (backlogged NewReno flows); and (iii) an equal mix of inelastic and elastic traffic. The duration of each experiment is 120 seconds.
We evaluate {\em accuracy}: the fraction of time \name{} correctly detects the presence of elastic cross-traffic. For each experiment, we report the mean accuracy of the detector across 5 runs. 

\cut{The key take-aways are:

\noindent\textbf{1) Cross-traffic RTT:} The elasticity detector observes the fluctuation in cross-traffic rate in the frequency domain, and thus doesn't need to know the cross-traffic RTT. In particular the cross-traffic RTT changes the phase but not the amplitude of the peak in the FFT.

\noindent\textbf{2) Mix of RTTs in the cross-traffic:} When the cross-traffic contains a mix of elastic flows with different RTTs, all elastic flows, regardless of the RTT, react to the elasticity detector's pulses, the superposition of their rates also oscillates, and the traffic is correctly classified as elastic.

\noindent\textbf{3) High cross-traffic load:} Using asymmetric pulses, a sender is able to create and observe fluctuations in the cross-traffic even when the sending rate is a small fraction of the link capacity.

\noindent\textbf{4) Comparison with Copa:} Compared to our elasticity detector, Copa mis-classifies cross-traffic when the inelastic traffic rate is high, or when elastic flows have high RTTs.}

\begin{figure}
    \includegraphics[width=\columnwidth]{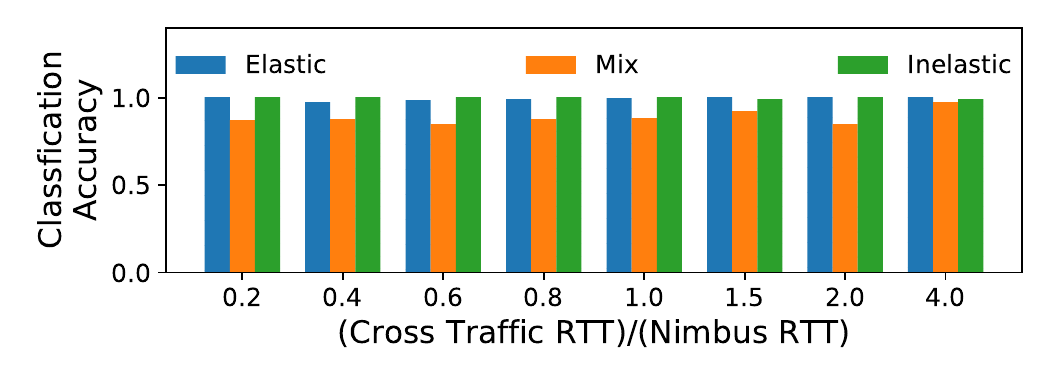}
    \vspace{-9mm}
    \caption{\small \name{} classifies purely elastic and inelastic traffic with accuracy greater than $98$\%. For a mix of elastic and inelastic traffic, the average accuracy is greater than $85$\% in all cases.}
    \label{fig:cross_rtt}
    \vspace{-5mm}
\end{figure}

\begin{figure}
    \includegraphics[width=\columnwidth]{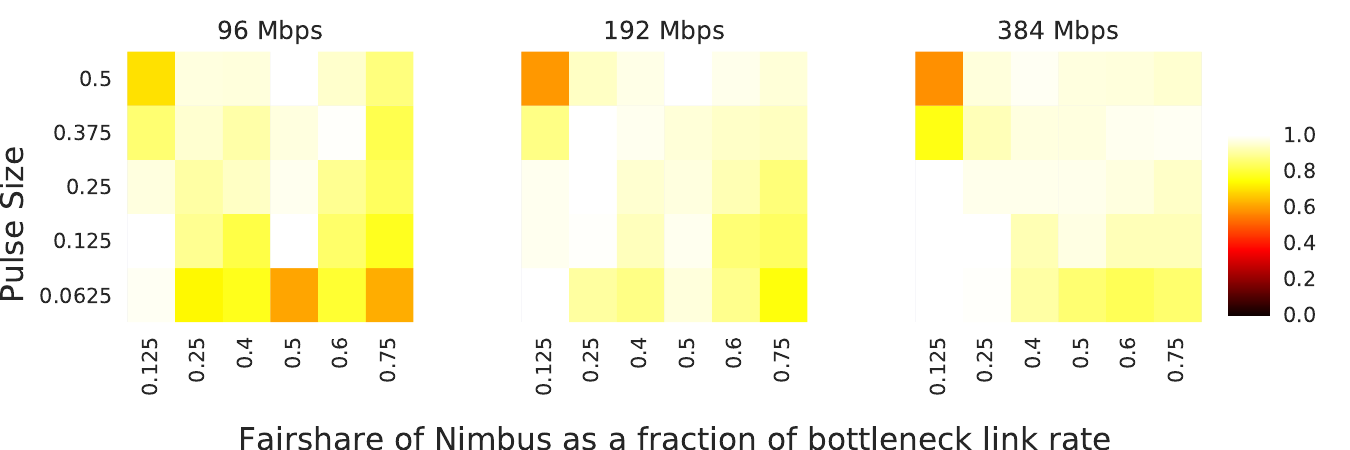}
    \vspace{-6mm}
    \caption{\small \name{} is robust to variations in link bandwidth and fraction of traffic controlled by it. The accuracy is high even when the fraction of traffic under control is small. Increasing pulse size increases robustness.}
    \label{fig:heatmap}
    \vspace{-4mm}
\end{figure}

\begin{figure}
    \centering
    \includegraphics[width=\columnwidth]{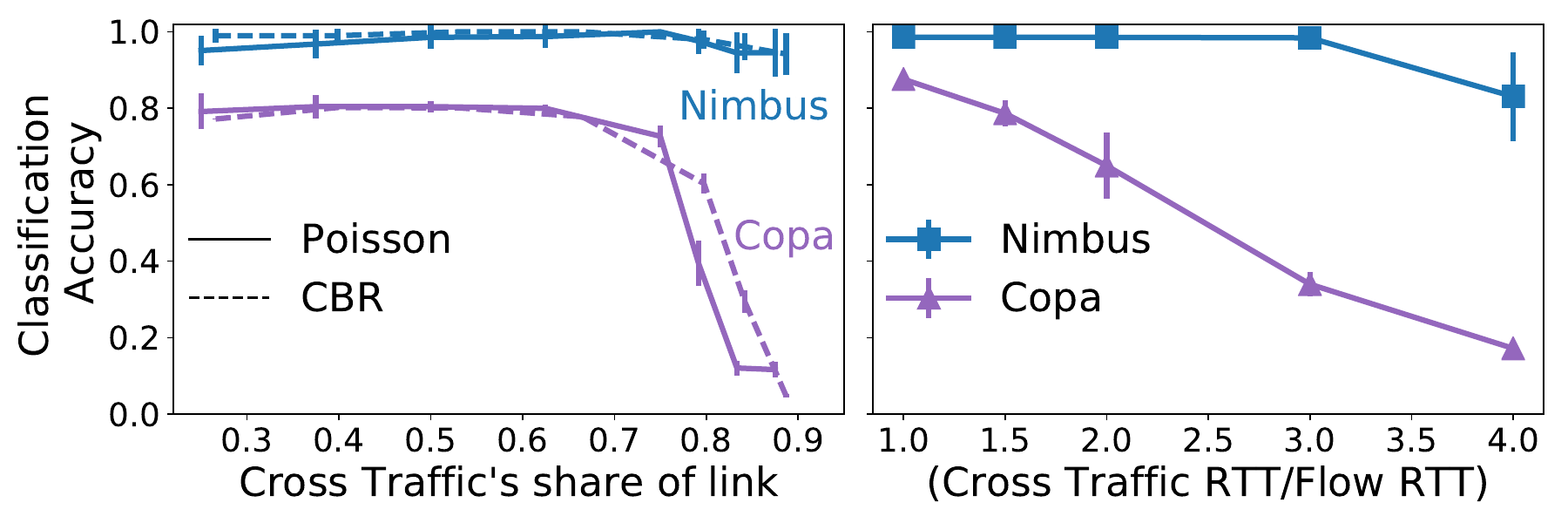}
    \vspace{-6mm}
    \caption{\small{\name{}'s classification accuracy is better \cut{has achieves higher classification accuracy}than Copa when (i) inelastic cross traffic occupies a large fraction of the link (left); (ii) elastic cross traffic has higher RTT than the flow's RTT (right). }}
    \vspace{-4mm}
    \label{fig:copa-compare}
\end{figure}

\smallskip
\noindent\textbf{Impact of cross-traffic \RTT.}
We vary the cross traffic's minimum RTT from 10 ms to 200 ms (0.2 -- 4$\times$ \name{}CC's RTT).
\Fig{cross_rtt} shows the mean detection accuracy for each of the three classes of cross traffic. We find that varying cross-traffic \RTT does not reduce accuracy. Regardless of the cross-traffic \RTT, the elastic flows respond to fluctuations created by \name{}, generating a peak in the cross-traffic FFT at the oscillation frequency. The cross traffic's \RTT \emph{affects the phase, but not the amplitude} of the peak in the FFT.

For purely inelastic and purely elastic traffic, the accuracy is more than 98\% in all cases, while for mixed traffic, the accuracy is more than 85\% in all cases (a random guess would have only achieved 50\%).  The accuracy for mixed traffic is lower because only half the cross traffic oscillates, and the FFT peak is smaller.

\smallskip
\noindent\textbf{A mix of \RTT{}s in the cross traffic.} We vary the number of elastic cross traffic flows from $1$ to $5$, where the \RTT of $n^{\text{th}}$ flow in $20 \cdot n$ ms. In case the cross-traffic contains elastic flows, all the elastic flows oscillate at \name{}'s pulse frequency. 
As a result, the sum of the rates of these elastic flows also oscillates,\footnote{Since the \RTT{}s are different, the elastic flows' oscillations will differ in phase and the oscillations could in theory cancel each other out leading to mis-classification, but it requires specific combinations of \RTT and is unlikely.} and the traffic is correctly classified as elastic. 
For purely elastic and inelastic traffic, \name{} achieves an average accuracy of $98$\% across 5 runs, while for mixed traffic, the mean accuracy is greater than $90$\% in all cases. In other words, heterogeneity in \RTT{}s of cross-flows does not degrade the accuracy of elasticity detection.

\smallskip
\noindent\textbf{Pulse size, link rate, and offered cross-traffic load.}
\label{s:multifactor-robustness-first-half}
We perform a multi-factor experiment varying \name{}'s pulse size from $1/16$ to $1/2$ the link rate, the fair share of the bottleneck link rate from 12.5\%---75\% (by varying the cross-traffic load), bottleneck link rates set to 96, 192, and 384 Mbit/s.
The accuracy for purely elastic cross-traffic is always higher than 95\%.
while the average accuracy over all the points for the other two traffic mixes is more than 90\%.
\Fig{heatmap} shows the average detection accuracy over the other two categories of cross-traffic (mix + purely inelastic).
%
%
The classification accuracy is not sensitive to cross traffic load. \name{}'s \emph{use of asymmetric pulses} enables a sender to create fluctuations in the cross traffic even when the sending rate is low. As a result, the detection accuracy remains high under high cross-traffic load. 

In general, increasing the pulse sizes improves accuracy because the elasticity detector can create a more easily observable change in the cross-traffic sending rates.
An increase in the link rate results in higher accuracy for a given pulse size and \name link share because the variance in the rates of inelastic Poisson cross-traffic reduces with increasing cross-traffic sending rate, reducing the number of false peaks in the cross-traffic FFT.
However, at low link rates, the elasticity detector has low accuracy ($\sim$60\%) when it uses high pulse sizes and controls a low fraction of the link rate. We believe that this is due to a quirk in the way the Linux networking stack reports round-trip time measurements under sudden sending rate changes.

\smallskip
\noindent\textbf{Impact of errors in link rate estimation.}
We explicitly supply an incorrect link rate estimate to \name. We vary the error in the link rate estimate from $-50$\% to $+50$\% of the real link rate value. The classification accuracy is high (> 80\%) for all traffic classes when the error is low ($\leq$ 12.5\%). When the error rate is higher, all traffic is classified as elastic. 

To understand why, define $\hat{z}(t)$ as the estimate of the cross traffic rate, $z^{*}(t)$ as the real cross traffic rate, $\hat{\mu}$ as the estimate of the link rate and $\mu^{*}$ as the real link rate. Then, from Equation ~\ref{eq:zfromR}:
\begin{equation}
    \hat{z}(t) = \hat{\mu} \frac{S(t)}{R(t)} - S(t),\qquad
      z^{*}(t) = \mu^{*} \frac{S(t)}{R(t)} - S(t) 
\vspace{-1mm}
\end{equation}
Combining the equations above, we get
\vspace{-2mm}
\begin{equation}
    \hat{z}(t) = \frac{\hat{\mu}}{\mu^{*}} z^{*}(t) + \big(\frac{\hat{\mu}}{\mu^{*}} - 1 \big) S(t) 
\end{equation}
\vspace{-4mm}

When the link estimate is inaccurate, the cross traffic estimate is a linear combination of the real cross traffic rate and the sending rate. As the error increases, the contribution of the sending rate to the cross traffic estimate increases. Since the sending rate oscillates at the pulse frequency, the cross traffic estimate also oscillates, and all cross traffic (regardless of its nature) is classified as elastic.

This implies that when the error in link rate estimate is high, \name{}CC always uses the TCP-competitive mode, losing the benefits of delay control in the presence of inelastic traffic. But its throughput will not suffer.

\smallskip
\noindent\textbf{Buffer size, \RTT{}, and Active Queue Management(AQM).} 
\label{s:multifactor-robustness-second-half}
\name{} is robust to these settings (\App{robustness:bufsize-rtt-aqm}).

\smallskip
\noindent\textbf{Comparison with Copa.}
\label{s:copa-compare} 
We now compare the classification accuracy of \name{} with Copa. 
First, we generate inelastic cross traffic at different rates and measure the accuracy.
We use a 96 Mbit/s bottleneck link with a 50 ms propagation delay and a 100 ms drop-tail buffer (2 BDP). We consider both constant-bit-rate (CBR) and Poisson cross traffic. \cut{(generated as described in $\S$\ref{s:big-experiment}).}

\Fig{copa-compare} (left) shows that \name{} has high accuracy in all cases, but Copa's accuracy drops sharply when the cross traffic occupies over 80\% of the link. 
This result highlights a pitfall of Copa's approach: setting an operating mode based on the absolute value of queueing delays is problematic.
With a high inelastic cross-traffic load, Copa is unable to drain the queue quickly enough (\ie every 5 RTTs), which throws off its detector.
In contrast, the elasticity detector estimates elasticity through delay variations caused by its pulses, and is more robust.

Next, we ran a backlogged \name{}CC or Copa flow competing against a backlogged NewReno flow. We vary the RTT of the NewReno flow between $1-4\times$ the RTT of the \name{}CC/Copa flow. \Fig{copa-compare} (right) shows that Copa's accuracy degrades as the RTT of the cross traffic increases; \name{}'s accuracy is much higher, dropping only slightly when the cross traffic RTT is $4\times$ larger than \name{}CC. 

An elastic cross-flow with a large RTT increases its rate slowly enough to evade detection by Copa. Therefore, Copa drains the queue as it expects and concludes the absence of non-Copa cross-traffic. This behavior continues until the cross-flow has grown to offer a load close to the link rate, when it starts interfering with Copa's queue draining. 
By contrast, \name{} is more robust since it is based on the time series of variations of the cross traffic rate\cut{queueing delay}. Moreover, even when the classification accuracy for Copa is higher, it makes frequent mode-switches and is suspectible to lose throughput against elastic traffic. \App{copa-compare} shows the throughput and queueing delay dynamics of Copa and \name{}CC.

\begin{figure}
    \centering
    \includegraphics[width=\columnwidth]{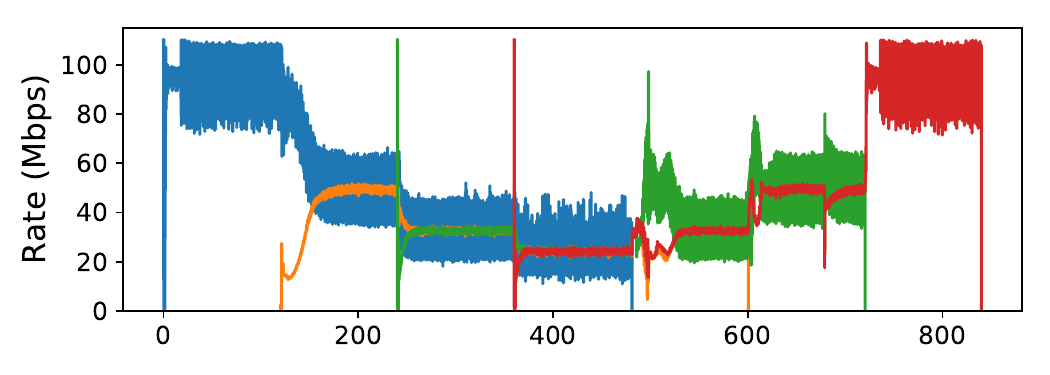}\vspace{-2mm}
    \includegraphics[width=\columnwidth]{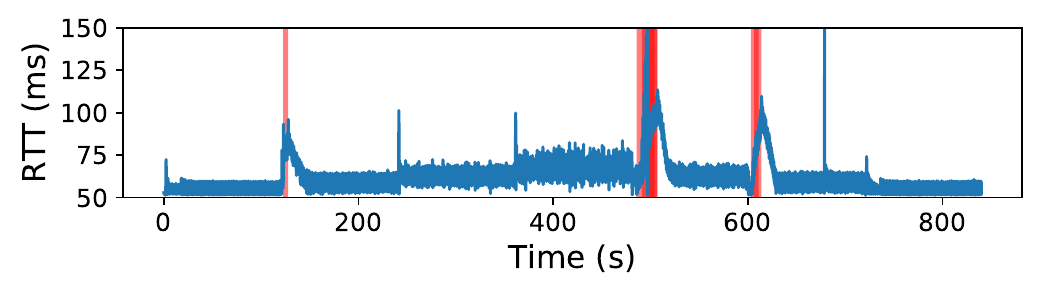}
    \vspace{-5mm}
    \caption{\small Multiple competing \name{}CC flows. Multiple \name{}CC flows achieve fair sharing of a bottleneck link (top graph). There is at most one pulser flow at any time; identified by its rate variations. Together, the flows achieve low delays by staying in delay mode for most of the duration (bottom graph). The red background shading shows when a \name{}CC flow was (incorrectly) in competitive mode}
    \label{fig:mul_staircase}
    \vspace{-2mm}
\end{figure}

\begin{figure}
    \centering
    \includegraphics[width=\columnwidth]{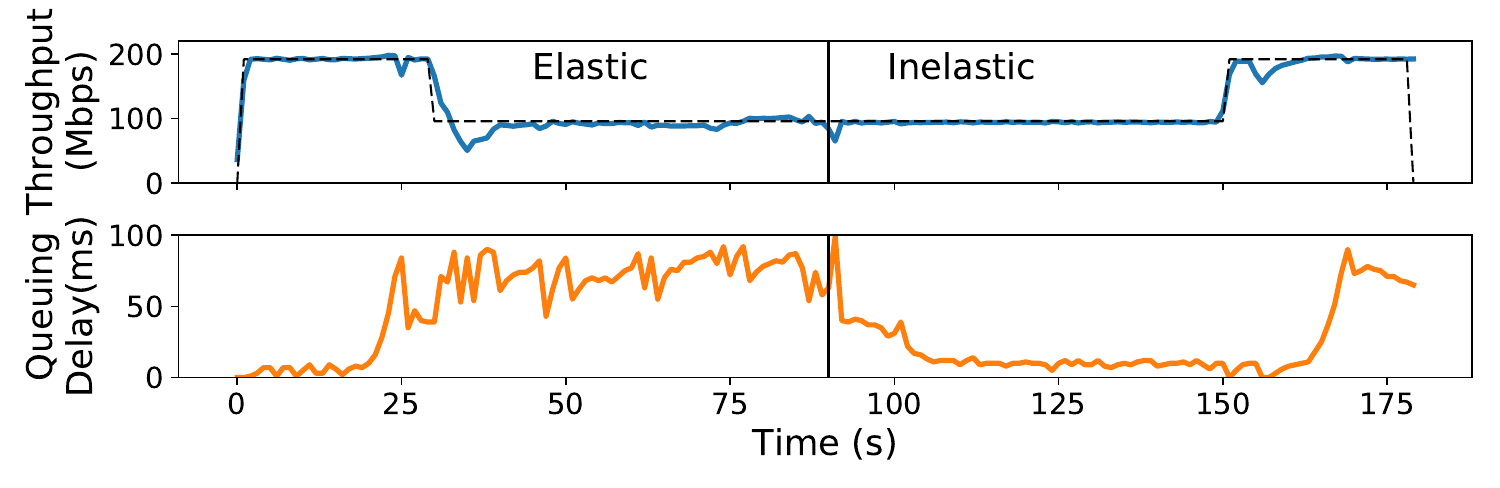}
    \vspace{-5mm}
    \caption{\small{{\bf Multiple \name{}CC flows  and other cross-traffic.} There are 3 \name{}CC flows throughout. Cross traffic in 30-90\,s is elastic and made up of three Cubic flows. Cross traffic in 90-150\,s is inelastic and made up of a 96 Mbit/s  constant bit-rate stream. \name{}CC flows achieve their fair share rate (top) while maintaining low delays in the absence of elastic cross traffic (bottom).}}
    \label{fig:mul_nimbus_switching}
    \vspace{-5mm}
\end{figure}

\begin{figure*}
    \centering
    \begin{subfigure}[b]{0.33\textwidth}
        \includegraphics[width=\textwidth]{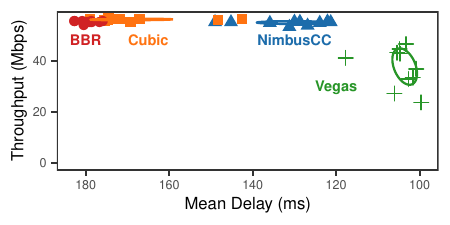}
        \vspace{-6mm}
        \caption{\small{EC2 California to Host A}}
        \label{fig:realworld:1}
    \end{subfigure}
    \begin{subfigure}[b]{0.33\textwidth}
        \includegraphics[width=\textwidth]{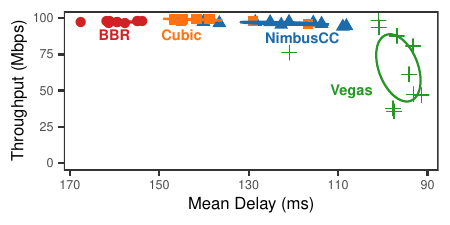}
        \vspace{-6mm}
        \caption{\small{EC2 Ireland to Host B}}
        \label{fig:realworld:2}
    \end{subfigure}
    \begin{subfigure}[b]{0.33\textwidth}
        \includegraphics[width=\textwidth]{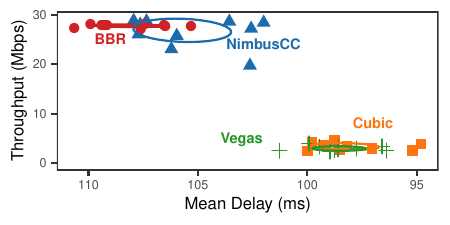}
        \vspace{-6mm}
        \caption{\small{EC2 London to Host C}}
        \label{fig:realworld:3}
    \end{subfigure}
    \vspace{-6mm}
    \caption{\small {{\bf Performance on three example Internet paths.} The $x$ axis is inverted; better performance is up and to the right. On paths with buffering and no drops, ((a) and (b)), \name{}CC achieves the same throughput as BBR and Cubic but reduces delays significantly. On paths with significant packet drops (c), Cubic suffers but \name{}CC achieves high throughput.}}
    \label{fig:realworld}
    \vspace{-4mm}
\end{figure*}

\begin{figure}
    \centering
    \begin{subfigure}[t]{0.45\textwidth}
        \includegraphics[width=\textwidth]{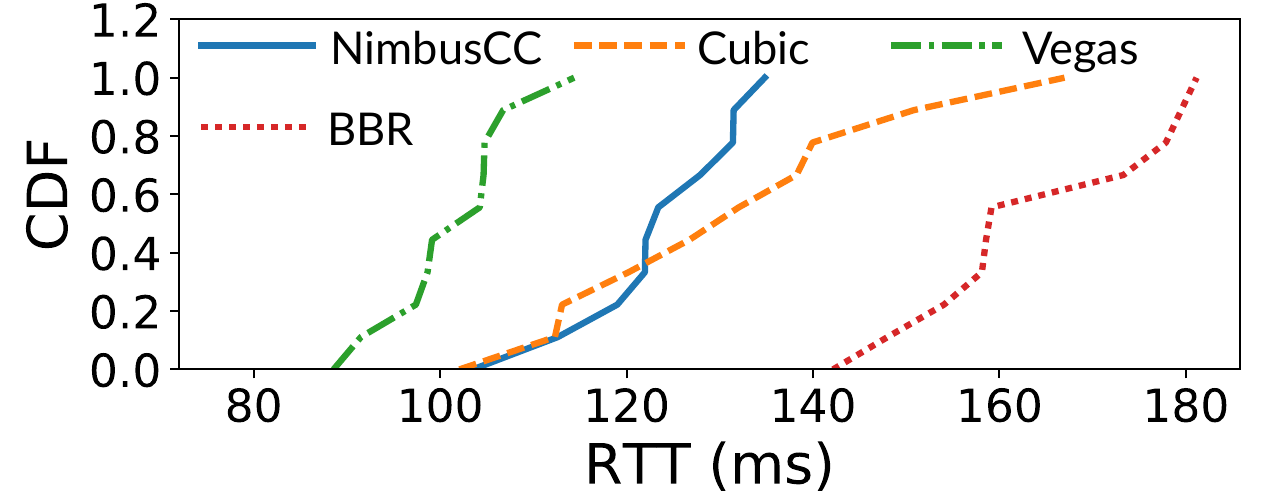}
        \label{fig:rw_cdf:delay}
    \vspace{-5mm}
    \end{subfigure}
    \begin{subfigure}[t]{0.45\textwidth}
        \includegraphics[width=\textwidth]{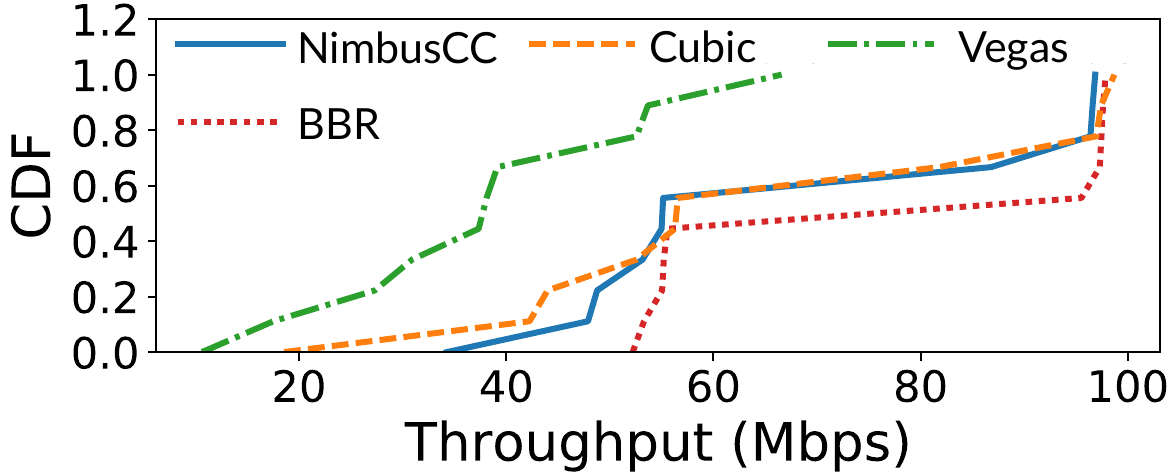}
        \label{fig:rw_cdf:rate}
    \end{subfigure}
    \vspace{-7mm}
    \caption{\small {{\bf Paths with queuing.} \name{}CC reduces the RTT compared to Cubic and BBR (40-50 ms lower), at similar throughput.}}
    \label{fig:rw_cdf}
    \vspace{-5mm}
\end{figure}

\subsection{Fairness With Elasticity Detection}
\label{s:multiple-nimbus-fair-sharing}

Can multiple flows run elasticity detection and share a bottleneck link fairly with each other and with cross-traffic?

We run \name{}CC with Vegas as its delay-control algorithm.
%
\Fig{mul_staircase} demonstrates how \name{}CC flows react as other \name{}CC flows arrive and leave (there is no other cross-traffic). Four flows arrive at a link with rate 96 Mbit/s and \rtt 50 ms. 
Each flow begins 120 s after the last one began, and lasts for 480 s. The top half shows the rates achieved by the four flows over time. Each new flow begins as a watcher. If the new flow detects a pulser ($t = 120, 240, 360$ s), it remains a watcher. If the pulser goes away or a new flow fails to detect a pulser, one of the watchers becomes a pulser ($t = 480, 720$ s). The pulser can be identified visually by its rate variations.

The flows share the link rate equally. The bottom half of the figure shows the achieved delays with red background shading to indicate when one of the flows is (incorrectly) in competitive-mode. The flows maintain low RTTs and stay in delay-mode for most of the time.

\Fig{mul_nimbus_switching} demonstrates multiple \name{}CC flows switching in the presence of cross-traffic. We run three \name{}CC flows on an emulated 192 Mbit/s link with a propagation delay of 50 ms. In the first 90\,s, the cross-traffic is elastic (three Cubic flows), and for the rest of the experiment, the cross-traffic is inelastic (96 Mbit/s constant bit-rate). The top graph shows the total rate of the three \name{}CC flows, along with a reference line for the fair-share rate of the aggregate. The graph at the bottom shows the measured queuing delays. 
\name{}CC shares the link fairly with other cross-traffic, and achieves low delays by staying in the delay mode in the absence of elastic cross-traffic.

\subsection{Performance on Internet Paths}
\label{s:realworld}


We ran \name{}CC on Internet paths on 25 paths between five senders and five receivers. The servers were Amazon EC2 instances located in California, London, Frankfurt, Ireland, and Paris, all with 10 Gbit/s links. The receivers were five residential hosts in different autonomous systems. We verified that the bottleneck in each case was not the server's Internet link.

To understand the nature of cross traffic on these paths, we ran experiments with \name{}CC delay-control algorithm (without mode-switching) and Cubic each performing bulk transfers over a three-day period. The results showed that scenarios where cross traffic is predominantly inelastic are common. 
This indicates that delay-control algorithms can be effective on the Internet (see \App{delay-control-motivation} for details).

%
To understand the nature of cross-traffic on these paths,
we initiated bulk data transfers using \name{}CC, Cubic, BBR, and Vegas.
We ran one-minute experiments over five hours on each path, and measured the achieved mean throughput and mean delay.
\Fig{realworld} shows throughput and delays over three of the paths. The $x$ (delay) axis is inverted; better performance is up and to the right. 
\name{}CC achieves high throughput comparable to BBR in all cases, at significantly lower delays. 
Cubic attains high throughput on paths with deep buffers (\Fig{realworld:1} and \Fig{realworld:2}), but not on paths with packet drops or policers (\Fig{realworld:3}).
%
Vegas attains poor throughput on these paths because it does not keep the bottleneck link busy and is unable to compete with elastic cross-traffic.
These trends show the utility of elasticity detection on Internet paths: it is possible to achieve high throughput and low delays over the Internet using delay-control algorithms with the ability to switch to a different competitive mode when required. 

\Fig{rw_cdf} summarizes the results on the paths with queueing. \name{}CC's throughput is similar to Cubic and $10$\% lower than BBR but at much lower delay (40--50 ms lower than BBR).
\label{p:end}
\newcommand{\conclsec}{Conclusion}
\section{\conclsec}
\label{s:concl}

This paper's key contribution is the idea that characterizing the {\em nature} of cross traffic is a useful signal and building block for congestion control. It introduced a method for detecting and quantifying the {\em elasticity} of cross traffic. The detection technique uses a carefully constructed asymmetric sinusoidal pulse and observes the frequency response of cross traffic rates at a sender, taking advantage of the property that elastic cross traffic can be made to oscillate at a pulsing frequency set by sender.  
We presented several experiments to demonstrate the robustness and accuracy of our proposed method. We also showed that elasticity detection enables transport protocols to combine the best aspects of delay-control methods while being competitive with buffer-filling flows when necessary.
We found that our proposed methods are beneficial not only on a variety of emulated conditions that model realistic workloads, but also on a collection of 25 real-world Internet paths.


\def\bibfont{\normalfont}
\bibliographystyle{abbrv}
\bibliography{nimbus}

\clearpage
\appendix

\section{Cross Traffic is Often Inelastic}
\label{app:delay-control-motivation}

\begin{figure}[tbh]
    \includegraphics[width=\columnwidth]{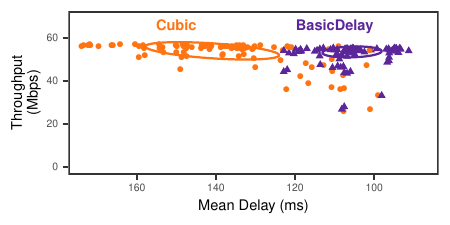}
    \vspace{-8mm}
    \caption{\small \textbf{Mean throughput and delay for 100 one-minute data transfers with Cubic and BasicDelay. The experiments were run between an AWS EC2 server in California and a receiver on the US east coast. The ellipses show one standard deviation. BasicDelay achieves the same throughput as Cubic in many runs, signifying an absence of elastic cross traffic in these cases.}}
    \label{fig:dsmotivation}
    \vspace*{-3pt}
\end{figure}

Our experiments on 25 Internet paths show that scenarios where cross traffic is predominantly inelastic are common. Figure~\ref{fig:dsmotivation} shows the average throughput and delay for 100 runs of buffer-filling Cubic compared to BasicDelay, a delay-controlling method, on one of these paths. The \dcc scheme generally achieves much lower delays than Cubic, with similar throughput. This shows that there is an opportunity to significantly improve delays using \dcc algorithms, provided we can detect the presence of elastic cross-traffic flows and compete with them fairly when needed. 

\section{\name{} Helps Cross Traffic}
\label{app:cross-traffic-impact}

\begin{figure}[t]
    \centering
    \includegraphics[width=0.48\textwidth]{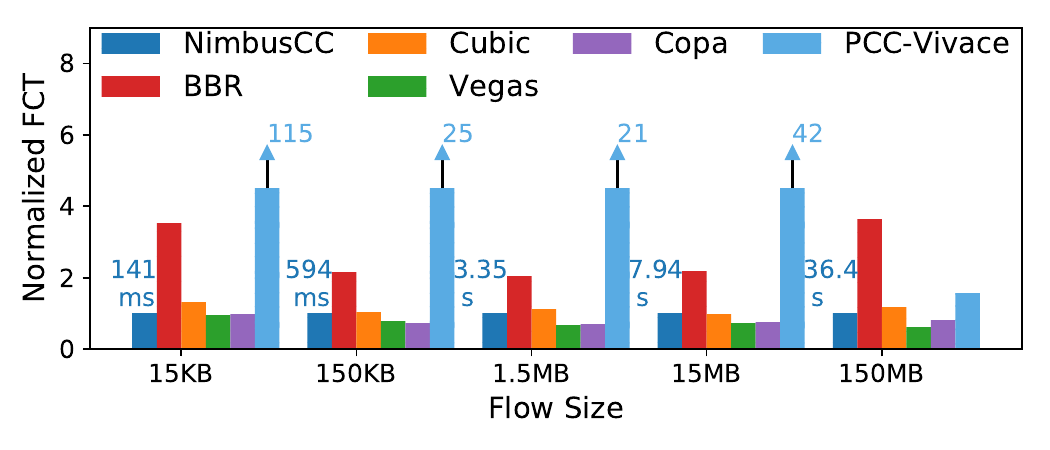}
    \vspace{-8mm}
    \caption{\small{Using \name{}CC reduces the p95 FCT of cross-flows relative to BBR at all flow sizes, and relative to Cubic for short flows. Vegas provides low cross-flow FCT, but its own rate is low.}}
    \label{fig:emp:fct}
\end{figure}

In the setup from \Sec{rw-wl}, we measure the flow completion time (FCT) of cross-traffic flows.
\Fig{emp:fct} compares the 95th percentile (p95) FCT for flows of different sizes. The FCTs are normalized by the corresponding value for \name{}CC at each flow size (i.e., \name{}CC is always 1).

BBR and PCC-Vivace exhibits much higher FCT at all cross-traffic flow sizes compared to the other protocols, consistent with the unfairness seen in the experiment in \S\ref{s:big-experiment}.

For small flows ($\leq$15 KB), the p95 FCT with \name{}CC and Copa are comparable to Vegas and lower than Cubic.
%
%
With \name{}CC, p95 FCT of cross traffic at higher flow sizes are slightly lower than Cubic because of small delays in switching to TCP-competitive mode.
At all flow sizes, Vegas provides the best cross-traffic flow FCTs, but its own flow rate is dismal; Copa is more aggressive than Vegas but less than \name{}CC, but at the expense of its own throughput  (\Sec{rw-wl}).

\section{\name{}CC \& Cubic v. BBR}
\label{app:bbr-comparison}
\begin{figure}[tbh]
    \includegraphics[width=\columnwidth]{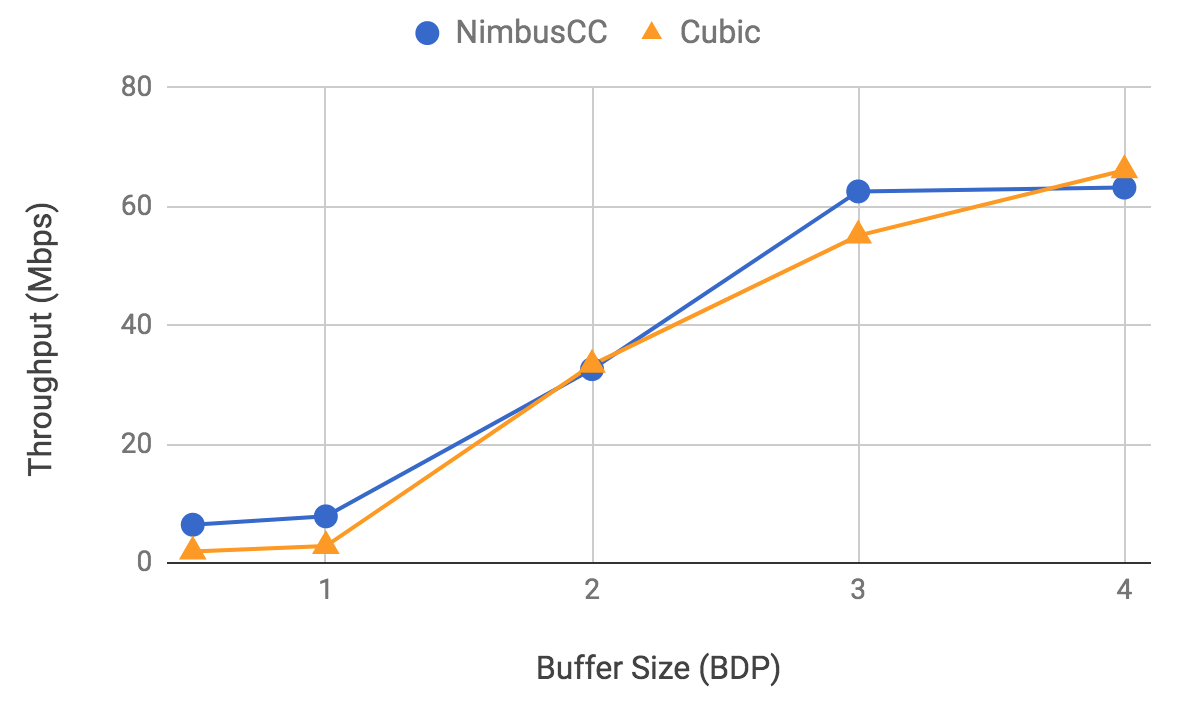}
    \vspace{-8mm}
    \caption{\textbf{ \name{}CC's performance against BBR is similar to that of Cubic.} Both \name{}CC and Cubic compete against 1 BBR flow on a 96 Mbit/s link. For various buffer sizes, \name{}CC achieves the same throughput as Cubic. }
    \label{fig:nimbus_vs_bbr}
\end{figure}
We now evaluate how well a \name{}CC (Cubic + BasicDelay) flow  competes with a BBR flow. 
In this experiment, the cross traffic is 1 BBR flow and the bottleneck link bandwidth is 96 Mbit/s. We vary the buffer size from 0.5 BDP to 4 BDP. \Fig{emp:switch} shows the mean throughput of \name{}CC and Cubic flows while competing with BBR over a 2-minute experiment. \name{}CC achieves the same throughput as Cubic for all buffer sizes.

In this experiment, when the buffer size is $\leq$ 1 BDP, BBR is not ACK-clocked, and the elasticity detector classifies it as inelastic traffic. As a result, \name{}CC gets a relatively small fraction of the link bandwidth. In this scenario, Cubic also gets a small fraction of the link, because BBR sends traffic at its estimate of bottleneck link and is too aggressive. 


When the buffer size is $\geq$1 BDP, BBR becomes ACK-clocked because of the cap on its congestion window. The elasticity detector now classifies BBR as elastic traffic. \name{}CC stays in competitive mode, so \name{}CC behaves like Cubic. 

\begin{figure}
    \centering
    \begin{subfigure}[b]{0.46\textwidth}
        \includegraphics[width=\textwidth]{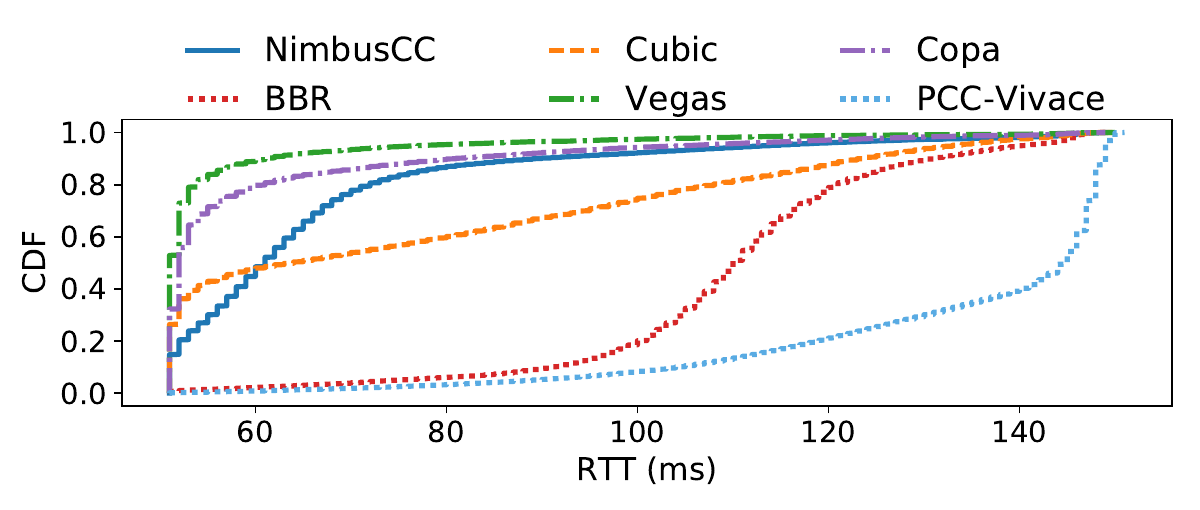}
        \vspace{-6mm}
        \caption{Per-packet RTT}
        \label{fig:emp_bbr:rtt}
    \end{subfigure}
    \begin{subfigure}[b]{0.46\textwidth}
        \includegraphics[width=\textwidth]{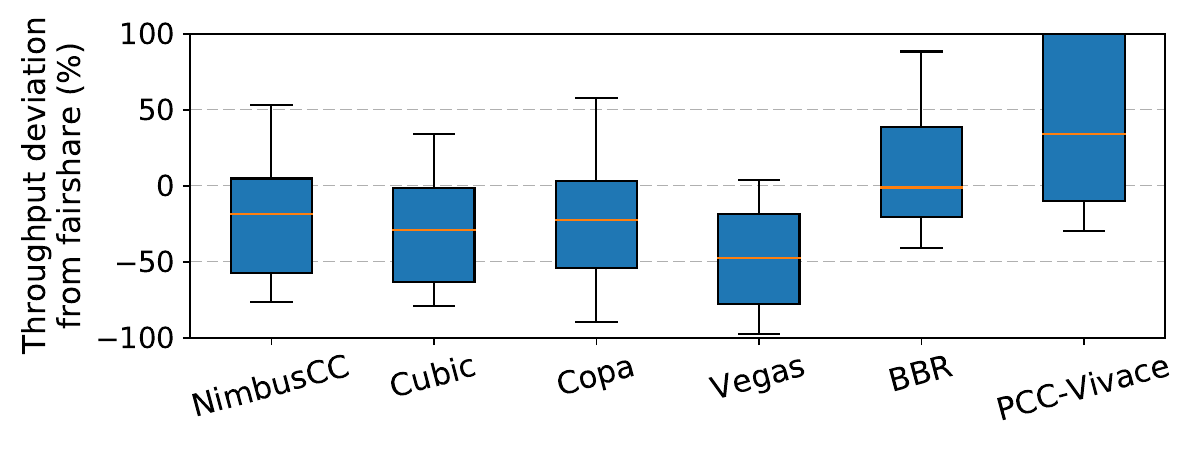}
        \vspace{-6mm}
        \caption{\small{Deviation from fairshare throughput}}
        \label{fig:emp_bbr:box}
    \end{subfigure}
    \vspace{-3mm}
    \caption{\small{{\bf WAN cross-traffic.} Cross traffic flows are running BBR. The deviation profile of \name{}CC is similar to that of Cubic, however, \name{}CC reduces delays.}}
    \label{fig:emp_bbr}
    \vspace{-4mm}
\end{figure}
\smallskip
\noindent\textbf{WAN-cross traffic:}
We repeated the experiment in \Fig{big}, with cross traffic flows running BBR. 
\Fig{emp_bbr} shows the performance of the various schemes. \name{}CC's deviation profile is similar to that of Cubic, but \name{}CC achieves lower delays. However, both \name{}CC and Cubic achieve lower throughput than the fairshare, this is because Cubic doesn't compete will with BBR cross traffic. Copa and Vegas have similar delays as \name{}CC, but their tail throughput is lower. PCC-Vivace is unfair to the cross traffic, and sends more than its fair-share.

\section{Copa's Mode-Switching Errors}
\label{app:copa-compare}

We explore the dynamics of \name{}CC and Copa's mode switching in experiments from the scenarios in $\S$\ref{s:copa-compare}. 

\subsection{CBR Cross Traffic}
\begin{figure}[t]
    \begin{center}
        \begin{subfigure}[b]{\columnwidth}
            \includegraphics[height=1.35in,width=0.8\columnwidth]{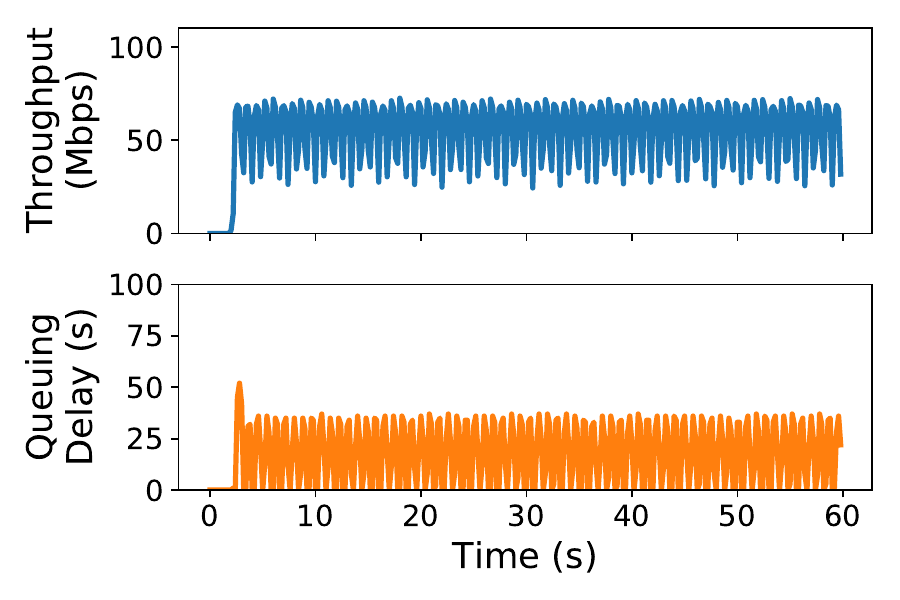}
            \vspace{-3mm}
            \subcaption{Copa: 24 Mbit/s CBR}
            \label{fig:appendix-cbr:copa-24}
        \end{subfigure}
        \begin{subfigure}[b]{\columnwidth}
            \includegraphics[height=1.35in,width=0.8\columnwidth]{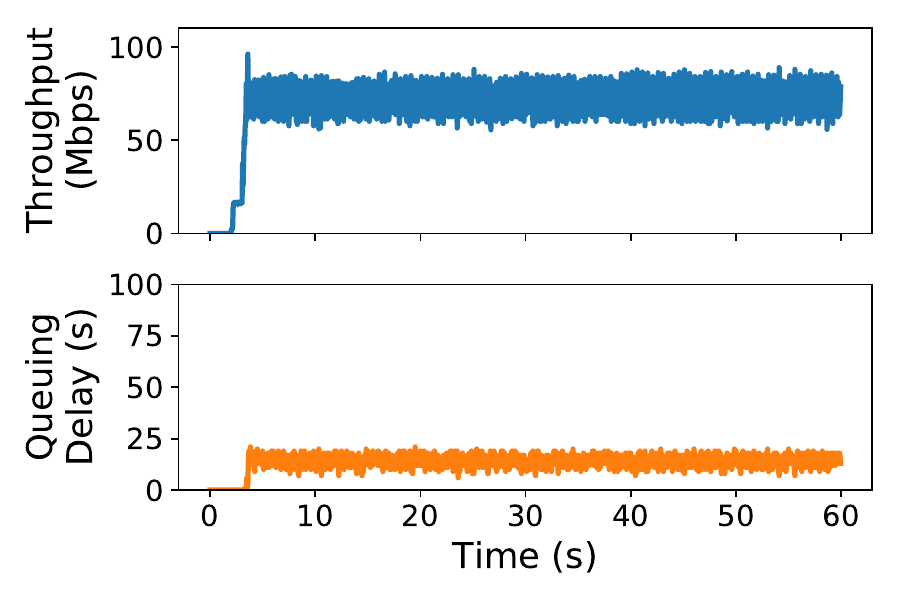}
            \vspace{-3mm}
            \subcaption{\name{}CC: 24 Mbit/s CBR}
            \label{fig:appendix-cbr:nimbus-24}
        \end{subfigure}
        \begin{subfigure}[b]{\columnwidth}
            \includegraphics[height=1.35in,width=0.8\columnwidth]{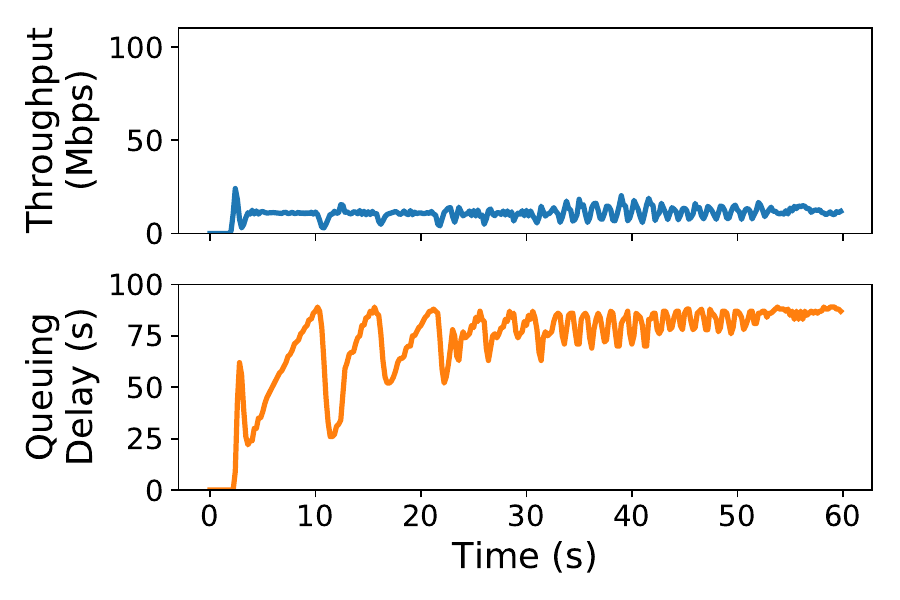}
            \vspace{-3mm}
            \subcaption{Copa: 80 Mbit/s CBR}
            \label{fig:appendix-cbr:copa-80}
        \end{subfigure}
        \begin{subfigure}[b]{\columnwidth}
            \includegraphics[height=1.35in,width=0.8\columnwidth]{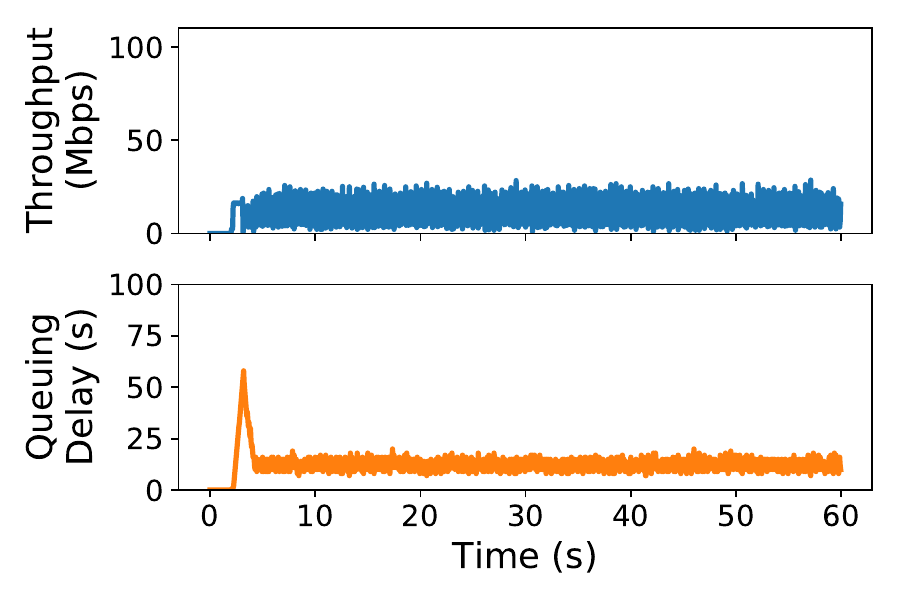}
            \vspace{-3mm}
            \subcaption{\name{}CC: 80 Mbit/s CBR}
            \label{fig:appendix-cbr:nimbus-80}
        \end{subfigure}
        \caption{When the CBR traffic is low (\subref{fig:appendix-cbr:copa-24}), Copa classifies the traffic as non buffer-filling and is able to achieve low queuing delays. But when the CBR traffic occupies a high fraction (\subref{fig:appendix-cbr:copa-80}), Copa incorrectly classifies the traffic as buffer-filling, resulting in higher queuing delays. In both the situations (\subref{fig:appendix-cbr:nimbus-24} and \subref{fig:appendix-cbr:nimbus-80}), the elasticity detector correctly classifies the traffic as inelastic and \name{}CC achieves low queuing delays.}
        \label{fig:appendix-cbr}
    \end{center}
\end{figure}

\Fig{appendix-cbr} shows throughput and delay profile for Copa and \name{}CC while competing against inelastic CBR traffic. We consider two scenarios: (i) CBR occupies a small fraction of the link (24 Mbits/s, 25\%) and (ii) CBR occupies majority of the link (80 Mbit/s, 83\%). 
When the CBR traffic is low (\Fig{appendix-cbr}\subref{fig:appendix-cbr:copa-24} and \Fig{appendix-cbr}\subref{fig:appendix-cbr:nimbus-24}), both Copa and Nimbus identify it as non-buffer-filling and inelastic, respectively, and achieve low queuing delays.

When the CBR's share of the link is high (\Fig{appendix-cbr}\subref{fig:appendix-cbr:copa-80}), Copa incorrectly classifies the cross traffic as buffer-filling and stays in competitive mode, leading to high queuing delays. Copa relies on a pattern of emptying queues to detect whether the cross traffic is buffer-filling or not. However, when the rate of cross traffic is $z$, the fastest possible rate at which the queue can drain is $\mu - z$, even if Copa reduces its rate to zero. If the cross traffic occupies $x$ fraction of the link (\ie $z = x \mu$), then
\begin{align}
    \max(-\frac{dQ}{dt}) &= \mu - z = (1 - x) \mu = (1 - x) \frac{BDP}{RTT}. 
\end{align}
Hence, if the queue size exceeds $5 \times  (1 - x) BDP$, Copa won't be able to drain the queue in 5 RTTs, and it will mis-classify the cross traffic as buffer-filling. The queue size can grow large due to a transient burst or if Copa incorrectly switches to competitive mode. Once Copa is in competitive mode, it will drive the queues higher, and may get stuck in that mode.

\name{} doesn't rely on emptying queues and correctly classifies cross traffic as inelastic, achieving low delays (\Fig{appendix-cbr}\subref{fig:appendix-cbr:nimbus-80}).

\subsection{Elastic cross traffic}
\begin{figure}
    \begin{center}
        \begin{subfigure}{\columnwidth}
            \includegraphics[height=1.35in,width=0.8\columnwidth]{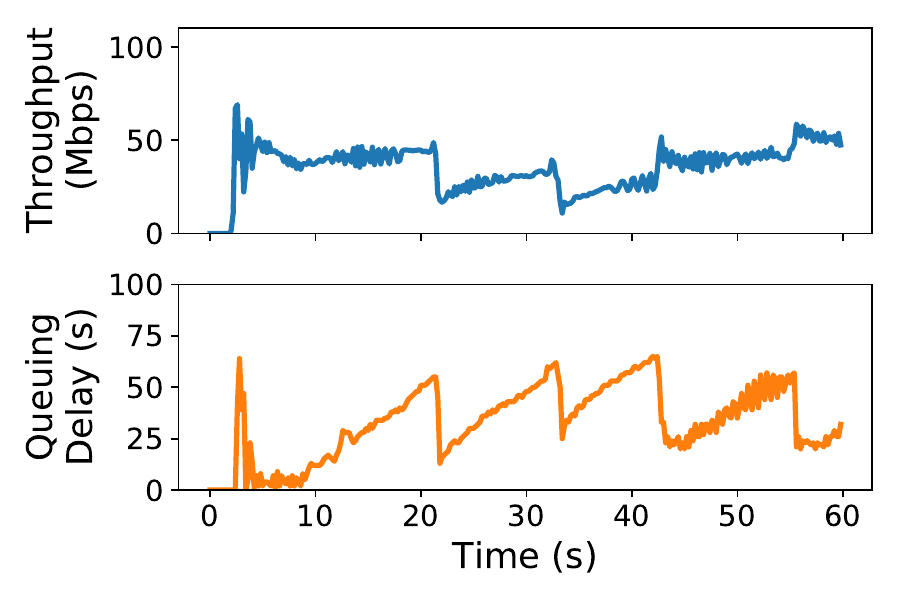}
            \vspace{-3mm}
            \subcaption{Copa: Cross Traffic RTT = 1 $\times$ Flow RTT}
            \label{fig:appendix-cross:copa-1}
        \end{subfigure}
        \begin{subfigure}{\columnwidth}
            \includegraphics[height=1.35in,width=0.8\columnwidth]{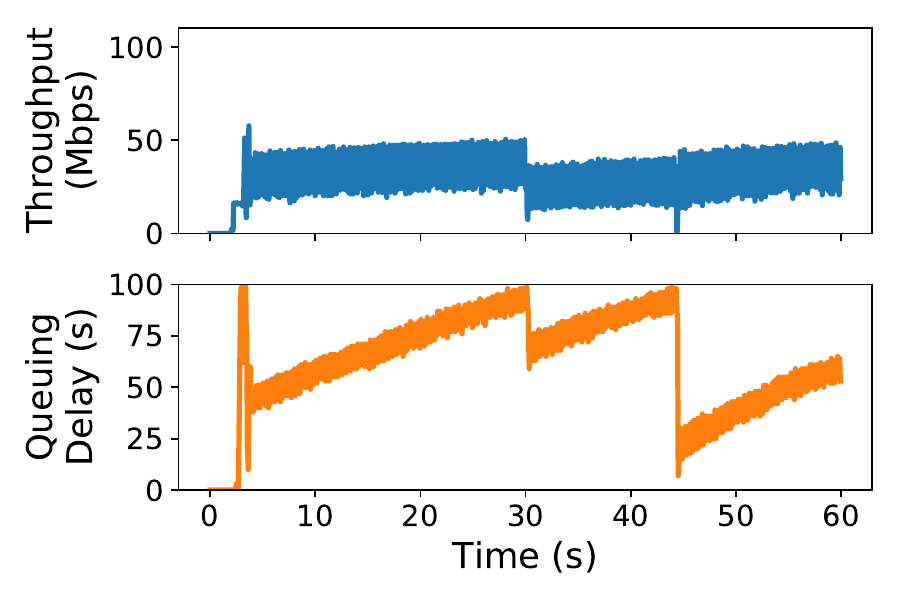}
            \vspace{-3mm}
            \subcaption{\name{}CC: Cross Traffic RTT = 1 $\times$ Flow RTT}
            \label{fig:appendix-cross:nimbus-1}
        \end{subfigure}
        \begin{subfigure}{\columnwidth}
            \includegraphics[height=1.35in,width=0.8\columnwidth]{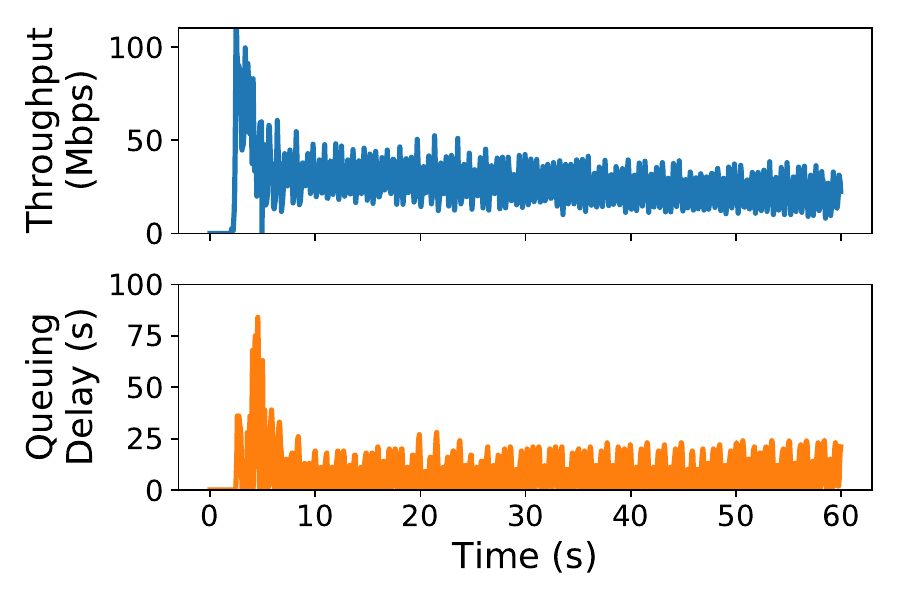}
            \vspace{-3mm}
            \subcaption{Copa: Cross Traffic RTT = 4 $\times$ Flow RTT}
            \label{fig:appendix-cross:copa-4}
        \end{subfigure}
        \begin{subfigure}{\columnwidth}
            \includegraphics[height=1.35in,width=0.8\columnwidth]{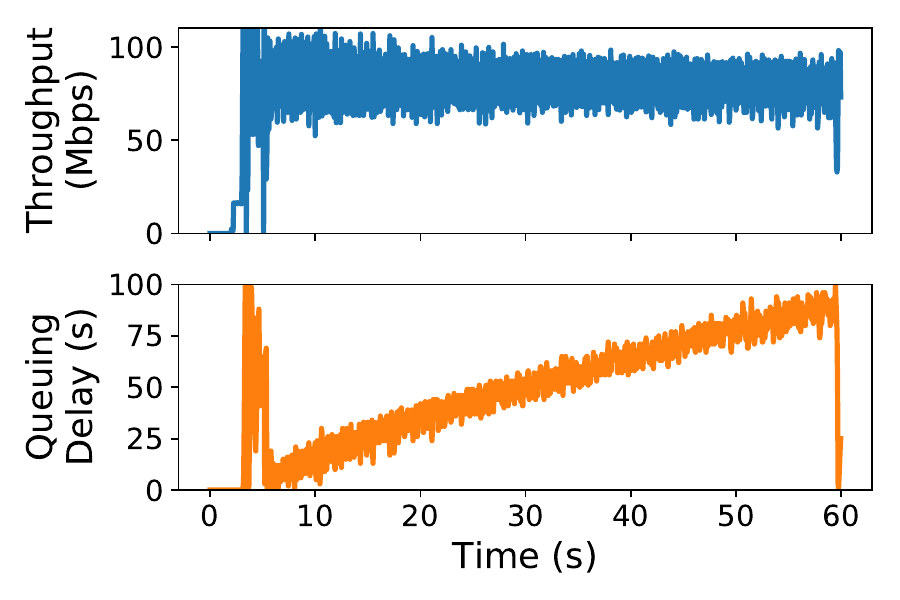}
            \vspace{-3mm}
            \subcaption{\name{}CC: Cross Traffic RTT = 4 $\times$ Flow RTT}
            \label{fig:appendix-cross:nimbus-4}
        \end{subfigure}
        \caption{\textbf{Queuing delay and throughput dynamics for elastic cross traffic.} When the elastic cross traffic increases fast enough (\subref{fig:appendix-cross:copa-1}), Copa classifies it as buffer-filling and is able to achieve its fair share. But when the elastic cross traffic increases slowly (\subref{fig:appendix-cross:copa-4}), Copa incorrectly classifies the traffic as non-buffer-filling, achieving less than its fair share. In both the situations (\subref{fig:appendix-cross:nimbus-1} and \subref{fig:appendix-cross:nimbus-4}), Nimbus correctly classifies the traffic as elastic and \name{}CC achieve its fair share. }
        \label{fig:appendix-cross}
    \end{center}
\end{figure}

\Fig{appendix-cross} shows throughput and delay over time for Copa and \name{}CC while competing against an elastic NewReno flow. We consider two scenarios: (1) both flows have the same propagation RTT, and (2) the cross traffic's propagation RTT is $4\times$ higher than the Copa or \name{}CC flow. When the RTTs are the same (\Fig{appendix-cross}\subref{fig:appendix-cross:copa-1} and \Fig{appendix-cross}\subref{fig:appendix-cross:nimbus-1}), both Copa and Nimbus correctly classify the cross traffic, achieving their fair share.

When the cross traffic RTT is higher (\Fig{appendix-cross}\subref{fig:appendix-cross:copa-4}), NewReno ramps up its rate slowly, causing Copa to mis-classify the traffic and achieve less than its fair share. Here, Copa achieves 27 Mbit/s but its fair share is at least 48 Mbit/s (in fact, 77 Mbit/s considering the RTT bias). In contrast, (\Fig{appendix-cross}\subref{fig:appendix-cross:nimbus-4}), Nimbus correctly classifies the cross traffic as elastic, and \name{}CC achieves its RTT-biased share of throughput. 

\section{Buffer size, RTT, and AQM}
\label{app:robustness:bufsize-rtt-aqm}
We vary the bottleneck drop-tail buffer size from 0.25 BDP to 4 BDP for three categories of cross traffic as in the earlier experiments, with propagation delays of 25 ms, 50 ms, and 75 ms. We also measured classification accuracy when the bottleneck link implements PIE~\cite{pie} at two target delays (0.25 BDP and 1 BDP) with a propagation delay of 50 ms.
With purely elastic or inelastic traffic, \name{} has a mean accuracy (across five runs) of 98\% or more in all cases but two, while with mixed traffic, the accuracy is always 85\% or more.
In all cases (including low accuracy ones), \name{}CC achieves its fair-share throughput and low delays.

Now we discuss the cases with low classification accuracy.
First, with shallow buffers of size less than the product of the delay threshold $x_{t}$ and the bottleneck link rate (\eg 0.25 BDP when the \rtt is 50 ms), \name{} classifies all traffic as elastic. Second, with the bottleneck link implementing PIE with small target delay (\eg corresponding to 0.25 BDP), \name{} classifies all traffic as elastic. In both cases, \name{}CC can incur heavy losses in delay-control mode as \name{}CC's target queuing delay of 0.25 BDP is comparable to the drop-tail buffer size or target delay of PIE. 
These losses interfere with the cross-traffic estimator leading to classification errors (in delay-control mode). However, low accuracy does not impact the performance of \name{}CC as it achieves its fair-share throughput and low delays (bounded by the small buffer size for a drop-tail queue and the delay control threshold of PIE).
Further, classification accuracy decreases when \name{}'s RTT exceeds its pulse period. Since \name{}'s measurements of rates are over one RTT, any oscillations over a smaller period cannot be observed.

\section{Elastic Flows, No ACK Clocking}
\label{app:slow-react}
\begin{figure}
    \centering
    \includegraphics[width=0.9\columnwidth]{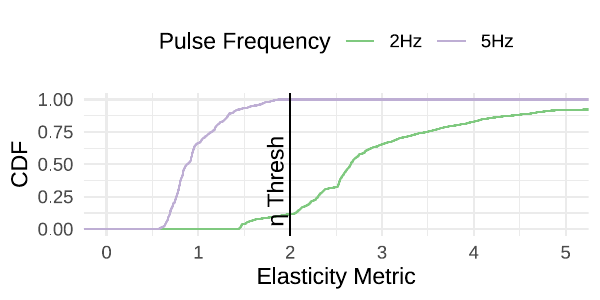}
    \caption{\small By modifying the pulse frequency, Nimbus correctly classifies PCC-Vivace, a rate-based elastic protocol, as elastic.}
    \label{fig:vivace-elastic}
\end{figure}

\name{} aims to detect ACK-clocked elastic flows that react quickly to changes in available bandwidth on RTT timescales. This experiment demonstrates \possname{} ability to also detect slow-reacting elastic cross traffic by tuning the pulse frequency. 
We ran a \name{}CC flow against a PCC-Vivace flow on a 96 Mbit/s link with 100ms of buffering. 
\Fig{vivace-elastic} shows the CDF of the elasticity metric, $\eta$, for two different pulse frequencies, $f_{p}$. 
PCC-Vivace is not ACK-clocked and does not react to Nimbus's pulses at $f_p =$ 5 Hz. 
As a result $\eta$ is below the threshold most of the time. 
Reducing the pulse frequency to 2 Hz creates pulses with a longer duration. 
PCC-Vivace reacts to these slower variations in available bandwidth, and is correctly classified as elastic ($\eta > \eta_{thresh}$). 

Changing the pulse frequency involves a trade-off. 
Increasing the pulse duration will increase queuing delays and congestion. 
But if slowly-reacting elastic protocols become widely deployed, competing with them using Nimbus for delay-control opportunities will require an increase in pulse duration.
\end{sloppypar}

\end{document}